\documentclass[twocolumn,superscriptaddress,floatfix,aps,prl]{revtex4-2}

\usepackage{graphicx}

\usepackage{amsmath,amssymb}
\usepackage{color,wrapfig,braket}
\usepackage{extarrows}
\usepackage{bm}

\usepackage[utf8]{inputenc}
\usepackage{natbib}
\usepackage{graphicx}
\usepackage{amsmath}
\usepackage{amssymb,bm,xspace,soul}
\usepackage{color}
\usepackage[dvipsnames]{xcolor}
\usepackage{anyfontsize} 
\usepackage{units}
\usepackage{times}
\usepackage{setspace}
\usepackage{braket}
\usepackage{empheq}
\usepackage{xcolor}
\usepackage{comment}

\usepackage{epsfig,psfrag,subfigure}
\usepackage{graphicx}
\usepackage{amsfonts}
\usepackage{exscale}
\usepackage{amsbsy}
\usepackage{stmaryrd}
\usepackage{wasysym}
\usepackage{verbatim} 
\usepackage{soul} 

\def\LSCO{La$_{2-x}$Sr$_x$CuO$_4$}
\def\LBCO{La$_{2-x}$Ba$_x$CuO$_4$}

\def\C60{A$_x$C$_{60}$}
\def\LNSCO{La$_{1.6-x}$Nd$_{0.4}$Sr$_x$CuO$_{4}$}

\def\LBCO{La$_{2-x}$Ba$_x$CuO$_4$}

\def\LNSCO{La$_{1.6-x}$Nd$_{0.4}$Sr$_x$CuO$_{4}$}

\def\HgCu3{HgCa$_2$Cu$_3$O$_{8+y}$}
\def\HgCu4{HgBa$_2$Ca$_3$Cu$_4$O$_{10+y}$}
\def\TlCu{Tl$_2$Ba$_2$CuO$_{6+\delta}$}
\def\TlCu3{Tl$_2$Ba$_2$Ca$_2$Cu$_3$O$_{10+y}$}
\def\TlCu4{Tl$_2$Ba$_2$Ca$_3$Cu$_4$O$_{12+y}$}

\def\BiCu3{Bi$_2$Sr$_2$Ca$_{2}$Cu$_3$O$_y$}

\def\8LSCO{La$_{1.88}$Sr$_{.12}$CuO$_4$}
\def\110LNSCO{La$_{1.5}$Nd$_{0.4}$Sr$_{0.1}$CuO$_{4}$}
\def\stage4LCO{La$_{2}$CuO$_{4+\delta}$}
\def\Y248{YBa$_2$Cu$_4$O$_8$}

\def\NbSe2{NbSe$_2$} 
\def\TaSe2{TaSe$_2$}
\def\TiSe2{TiSe$_2$}
\def\NaCoOH2O{Na$_{0.3}$CoO$_{2y}$H$_2$O}
\def\MgB2{MgB${}_2$}

\begin{document}

\title{Generic character of charge and spin density waves in superconducting cuprates}

\author{Sangjun Lee}
\affiliation{Department of Physics, University of Illinois at Urbana-Champaign, Urbana, Illinois 61801, USA}
\affiliation{Materials Research Laboratory, University of Illinois at Urbana-Champaign, Urbana, Illinois 61801, USA}
\author{Edwin W. Huang} 
\affiliation{Department of Physics, University of Illinois at Urbana-Champaign, Urbana, Illinois 61801, USA}
\affiliation{Institute of Condensed Matter Theory, University of Illinois, Urbana, Illinois 61801, USA}
\author{Thomas A. Johnson}
\affiliation{Department of Physics, University of Illinois at Urbana-Champaign, Urbana, Illinois 61801, USA}
\affiliation{Materials Research Laboratory, University of Illinois at Urbana-Champaign, Urbana, Illinois 61801, USA}
\author{Xuefei Guo}
\affiliation{Department of Physics, University of Illinois at Urbana-Champaign, Urbana, Illinois 61801, USA}
\affiliation{Materials Research Laboratory, University of Illinois at Urbana-Champaign, Urbana, Illinois 61801, USA}
\author{Ali A. Husain}
\affiliation{Department of Physics, University of Illinois at Urbana-Champaign, Urbana, Illinois 61801, USA}
\affiliation{Materials Research Laboratory, University of Illinois at Urbana-Champaign, Urbana, Illinois 61801, USA}
\author{Matteo Mitrano}
\affiliation{Department of Physics, University of Illinois at Urbana-Champaign, Urbana, Illinois 61801, USA}
\affiliation{Materials Research Laboratory, University of Illinois at Urbana-Champaign, Urbana, Illinois 61801, USA}
\author{Kannan Lu}
\affiliation{Department of Physics, University of Illinois at Urbana-Champaign, Urbana, Illinois 61801, USA}
\affiliation{Materials Research Laboratory, University of Illinois at Urbana-Champaign, Urbana, Illinois 61801, USA}
\author{Alexander V. Zakrzewski}
\affiliation{Department of Physics, University of Illinois at Urbana-Champaign, Urbana, Illinois 61801, USA}
\affiliation{Materials Research Laboratory, University of Illinois at Urbana-Champaign, Urbana, Illinois 61801, USA}
\author{Gilberto A. de la Pe{\~n}a}
\affiliation{Department of Physics, University of Illinois at Urbana-Champaign, Urbana, Illinois 61801, USA}
\affiliation{Materials Research Laboratory, University of Illinois at Urbana-Champaign, Urbana, Illinois 61801, USA}
\author{Yingying Peng}
\affiliation{Department of Physics, University of Illinois at Urbana-Champaign, Urbana, Illinois 61801, USA}
\affiliation{Materials Research Laboratory, University of Illinois at Urbana-Champaign, Urbana, Illinois 61801, USA}
\author{Hai Huang}
\affiliation{Stanford Synchrotron Radiation Lightsource, SLAC National Accelerator Laboratory, Menlo Park, California 94025, USA}
\author{Sang-Jun Lee}
\affiliation{Stanford Synchrotron Radiation Lightsource, SLAC National Accelerator Laboratory, Menlo Park, California 94025, USA}
\author{Hoyoung Jang}
\affiliation{Stanford Synchrotron Radiation Lightsource, SLAC National Accelerator Laboratory, Menlo Park, California 94025, USA}
\author{Jun-Sik Lee}
\affiliation{Stanford Synchrotron Radiation Lightsource, SLAC National Accelerator Laboratory, Menlo Park, California 94025, USA}
\author{Young Il Joe}
\affiliation{National Institute of Standards and Technology, Boulder, Colorado 80305, USA}
\author{William B. Doriese}
\affiliation{National Institute of Standards and Technology, Boulder, Colorado 80305, USA}
\author{Paul Szypryt}
\affiliation{National Institute of Standards and Technology, Boulder, Colorado 80305, USA}
\author{Daniel S. Swetz}
\affiliation{National Institute of Standards and Technology, Boulder, Colorado 80305, USA}
\author{Songxue Chi}
\affiliation{Neutron Scattering Division, Oak Ridge National Laboratory, Oak Ridge, Tennessee 37831, USA}
\author{Adam A. Aczel}
\affiliation{Neutron Scattering Division, Oak Ridge National Laboratory, Oak Ridge, Tennessee 37831, USA}
\author{Gregory J. MacDougall}
\affiliation{Department of Physics, University of Illinois at Urbana-Champaign, Urbana, Illinois 61801, USA}
\affiliation{Materials Research Laboratory, University of Illinois at Urbana-Champaign, Urbana, Illinois 61801, USA}
\author{Steven A. Kivelson}
\affiliation{Department of Physics, Stanford University, Stanford, California 94305, USA}
\author{Eduardo Fradkin}
\affiliation{Department of Physics, University of Illinois at Urbana-Champaign, Urbana, Illinois 61801, USA}
\affiliation{Institute of Condensed Matter Theory, University of Illinois, Urbana, Illinois 61801, USA}
\author{Peter Abbamonte}
\affiliation{Department of Physics, University of Illinois at Urbana-Champaign, Urbana, Illinois 61801, USA}
\affiliation{Materials Research Laboratory, University of Illinois at Urbana-Champaign, Urbana, Illinois 61801, USA}

\begin{abstract} 
Understanding the nature of charge density waves (CDW) in cuprate superconductors has been complicated by material specific differences. A striking example is the opposite doping dependence of the CDW ordering wavevector in La-based and Y-based compounds, the two families where charge ordering is strongest and best characterized. 
Here we report a combined resonant soft X-ray scattering (RSXS) and neutron scattering study of charge and spin density waves in isotopically enriched La$_{1.8-x}$Eu$_{0.2}$Sr$_{x}$CuO$_{4}$ over a range of doping $0.07 \leq x \leq 0.20$. For all dopings studied by RSXS, we find  
that the CDW amplitude is approximately temperature-independent and develops well above experimentally accessible temperatures. Surprisingly, the CDW ordering wavevector shows a non-monotonic temperature dependence, with a sudden change occurring at temperatures near the SDW onset temperature. We describe this behavior with a Landau-Ginzburg theory for an incommensurate CDW in a metallic system with a finite charge compressibility and CDW-SDW coupling. 
Our Landau-Ginzburg analysis suggests that the ordering wavevector at high temperatures decreases with increased doping. This behavior is opposite to the trend at low temperatures and highly reminiscent of the doping dependence seen in YBa$_2$Cu$_3$O$_{6+\delta}$, suggesting a common origin of the CDW in hole-doped cuprate superconductors.
\end{abstract}

\date{\today}

\maketitle

Charge density waves (CDWs) are pervasive in cuprate superconductors and are now believed to be fundamental to the low temperature properties of these materials. A perplexing issue has been that the evolution of the wave vector of the CDW, $Q_\mathrm{CDW}$, upon doping has different trends in different cuprates \cite{CominReview2016}. In YBa$_{2}$Cu$_{3}$O$_{6+\delta}$ (YBCO) \cite{Ghiringhelli2012,Blackburn2013,Hucker2014} and Bi-based cuprates \cite{Comin2014, Fujita2012, daSilvaNeto2014} $Q_\mathrm{CDW}$ decreases with increasing doping, while La-based cuprates \cite{Abbamonte2005, Hucker2011, Croft2014} exhibit the so-called Yamada relationship \cite{Yamada1998} in which $Q_\mathrm{CDW}$ increases with doping, $x$, as $Q_\mathrm{CDW}=2x$ for $x\leq0.125$ and saturates at $Q_\mathrm{CDW}=0.25$ for $x>0.125$. This disparity raises the question of whether the CDWs in different cuprates are fundamentally different, or if there is an underlying generic form of CDW order that simply manifests in different ways for material specific reasons.

The $Q_\mathrm{CDW}$ at a given doping and temperature, $T$, may be influenced by many factors, such as pinning on the lattice or other types of order that may also be present \cite{Dean2017,Nie2017,Dean2019}. Such extrinsic effects may cause $Q_\mathrm{CDW}$ to shift away from the ``natural'' ordering wave vector that the system intrinsically favors. Thus, it is essential to identify and disentangle the various effects that interact with the CDW. In particular, a spin density wave (SDW) is present in La-cuprates that is absent in YBCO and Bi-cuprates, suggesting that the presence of spin order may be a key factor in the distinct behaviors of $Q_\mathrm{CDW}$ in these materials \cite{Dean2017, Nie2017,Dean2019}. A close relationship between SDW and CDW order was previously observed in stripe-ordered nickelates \cite{Tokura2001} where $Q_\mathrm{CDW}$ is influenced by commensuration with the lattice and pinning with the SDW. 
What enabled these effects to be disentangled was a full study of the doping and temperature dependence, $Q_\mathrm{CDW}(x, T)$ \cite{Tokura2001}, and a similar strategy could shed light on the cuprates.
The La-cuprates are in some ways ideal for this; they have strong CDW and SDW tendencies, the concentration of doped hole is readily determined, and it can be varied across the entire range of the superconducting dome.  However, they have the complication of a transition to the low-temperature tetragonal (LTT) structure at a similar temperature as the onset of CDW order.

Here, using resonant soft x-ray scattering (RSXS) and neutron scattering, we present a study of the CDW and SDW in La$_{1.8-x}$Eu$_{0.2}$Sr$_{x}$CuO$_{4}$ (LESCO) over a wide range of doping $0.07 \leq x \leq 0.20$ and temperatures up to 270 K. We used isotopically enriched LESCO crystals with $^{153}$Eu for the neutron study in order to reduce the neutron absorption cross section, giving us improved sensitivity to the SDW (see {\it Materials and Methods}). LESCO has the advantage over other La-cuprates that the LTT transition temperature is almost fixed to $T_\mathrm{LTT} \sim 130$ K over the whole phase diagram \cite{Klauss2000, Fujita2002}. Hence, while the LTT distortion may influence the CDW, it at least does not change with doping, simplifying the interpretation. RSXS measurements were performed on LESCO crystals with $x=0.07$, 0.10, 0.125, 0.15, 0.17, 0.20 and the CDW was observed in all dopings except for $x=0.07$. Neutron scattering measurements were performed on $x=0.07$, 0.10, 0.11, 0.125, 0.15 and the SDW was observed in all dopings. 

We find, at low temperature, that the SDW and CDW in LESCO exhibit the expected Yamada behavior \cite{Yamada1998}, consistent with previous findings for the CDW \cite{Fink2011}. At higher temperatures, however, the CDW exhibits two distinct behaviors. First, the CDW does not show a clear phase transition. As previously demonstrated in Ref. \cite{Chang2020} for $x=0.125$, while the CDW peak intensity decreases with temperature, its integrated intensity is constant over the entire temperature range measured. We find that this behavior persists for all dopings, implying that the CDW amplitude is approximately temperature-independent everywhere in the phase diagram where the CDW is present. Using a non-linear sigma model, we show that this behavior is consistent with a CDW with very high mean-field transition temperature (far above room temperature) in the presence of weak, quenched disorder that induces phase fluctuations without significantly altering the CDW amplitude.

Second, we find that $Q_\mathrm{CDW}$ exhibits a complicated, non-monotonic dependence on doping and temperature, consistent with a competition between multiple effects. We explain this behavior quantitatively using a Landau theory with a new effect, the compressibility of the uncondensed metallic electrons \cite{Brown-2005}, which competes with the natural $Q_\mathrm{CDW}$ as well as its pinning on the SDW. 
In the context of this theory, the experimental results can be extrapolated to high temperatures where  extrinsic effects are minimal, where an ``intrinsic'' $Q_\mathrm{CDW}$ can be identified that decreases with increasing $x$, and is even quantitatively similar to the ordering vectors observed in YBCO and Bi-cuprates. This finding suggests that the CDW physics is generic across all hole doped cuprates, although its low temperature manifestations differ due to material-specific details, and moreover that it is characterized by higher temperature/energy scales than has been previously appreciated.

\begin{figure}[hbt]
\centering
\includegraphics[width=\linewidth]{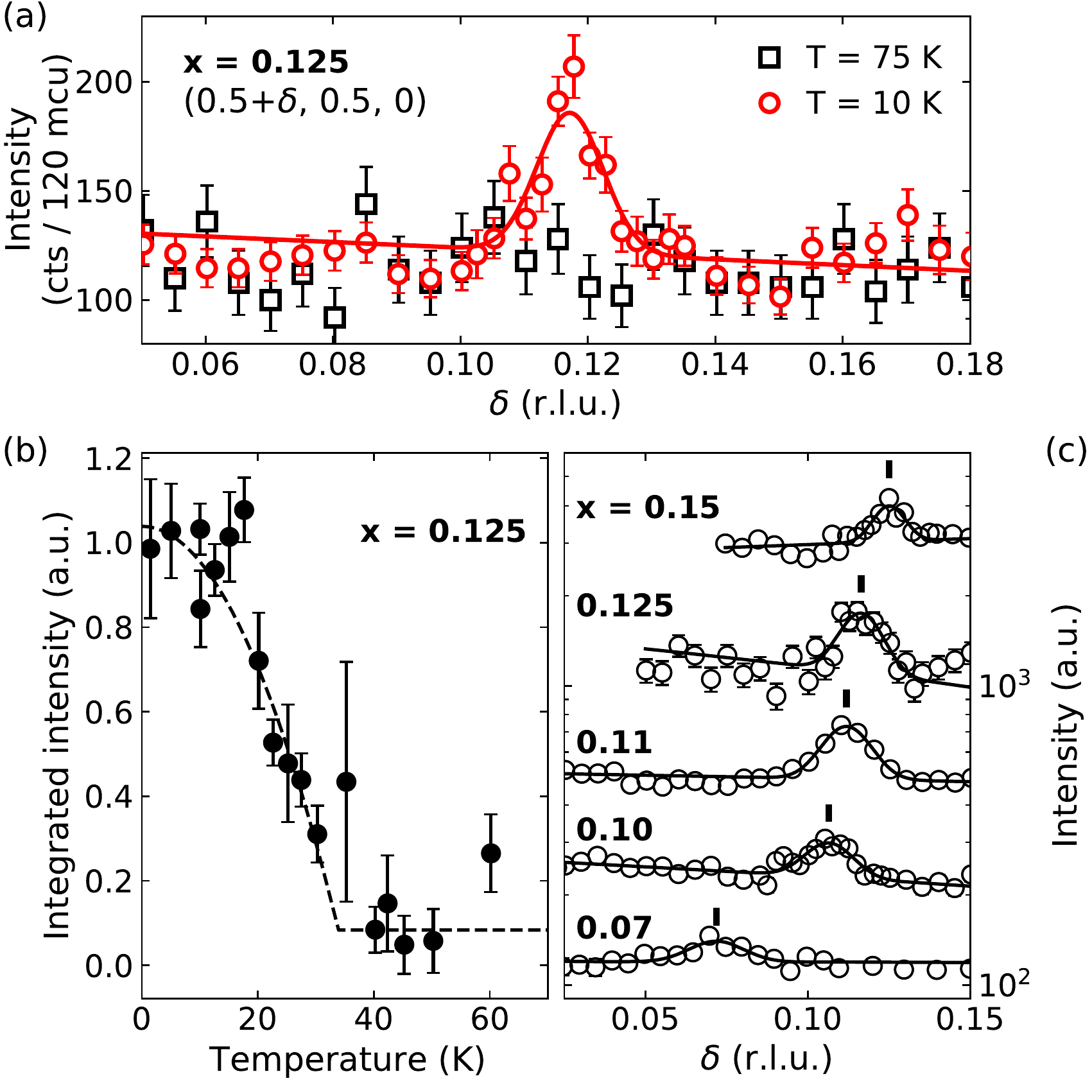}
\caption{(a) $H$-scans of SDW in LESCO $x = 0.125$ at $T=10$ K, and 75 K at which the SDW peak is absent. The scans are plotted against $\delta$ where $H=0.5+\delta$. The solid line is a fit to a Gaussian function. (b) Integrated intensity of SDW, inferred from fits to Gaussian functions. The dashed line shows a fit to a square-root model (described in the text) to determine SDW onset temperatures. Error bars correspond to standard errors. (c) $H$-scans of SDW measured at 4 K for $x$ = 0.11, and 1.5 K for all other dopings. The solid lines correspond to a fit with Gaussian peaks on a linear background. The black ticks above each curve indicate the center positions of the peaks.}
\label{fig:fig1}
\end{figure}

We first show our neutron scattering results in Fig. \ref{fig:fig1}. Fig. \ref{fig:fig1}(a) shows elastic momentum scans of the SDW in LESCO $x=0.125$ (see also {\it SI Appendix III} for the full scan covering negative and positive range of $\delta$). For all samples measured with neutron scattering, we observe SDW peaks that disappear with increasing temperature. We fit these peaks to Gaussian functions, and show the integrated intensity for LESCO $x=0.125$ in Fig. \ref{fig:fig1}(b). To define an onset temperatures, we fit the function $I(T) = I_0[1 - (T / T_{\mathrm{SDW}})^{2}] + C$ (for $T\leq T_{\mathrm{SDW}}$) to the integrated intensities and show the inferred onset temperatures in Fig. \ref{fig:fig4}. We make two key observations. First, the inferred wavevector is temperature-independent with a doping dependence following the well-known Yamada relation, where Q$_{\mathrm{SDW}}$ $\sim$ $x$ until saturation above $x = 0.125$ \cite{Yamada1998}, as shown in Fig. \ref{fig:fig6}. Second, we observe an onset temperature of 10 K for $x=0.07$, and near 40 K for the other dopings (see {\it SI Appendix III}), which is consistent with the neutron scattering result for x=0.15 from Ref. \cite{Hucker2007}.

\begin{figure}[hbt]
\centering
\includegraphics[width=\linewidth]{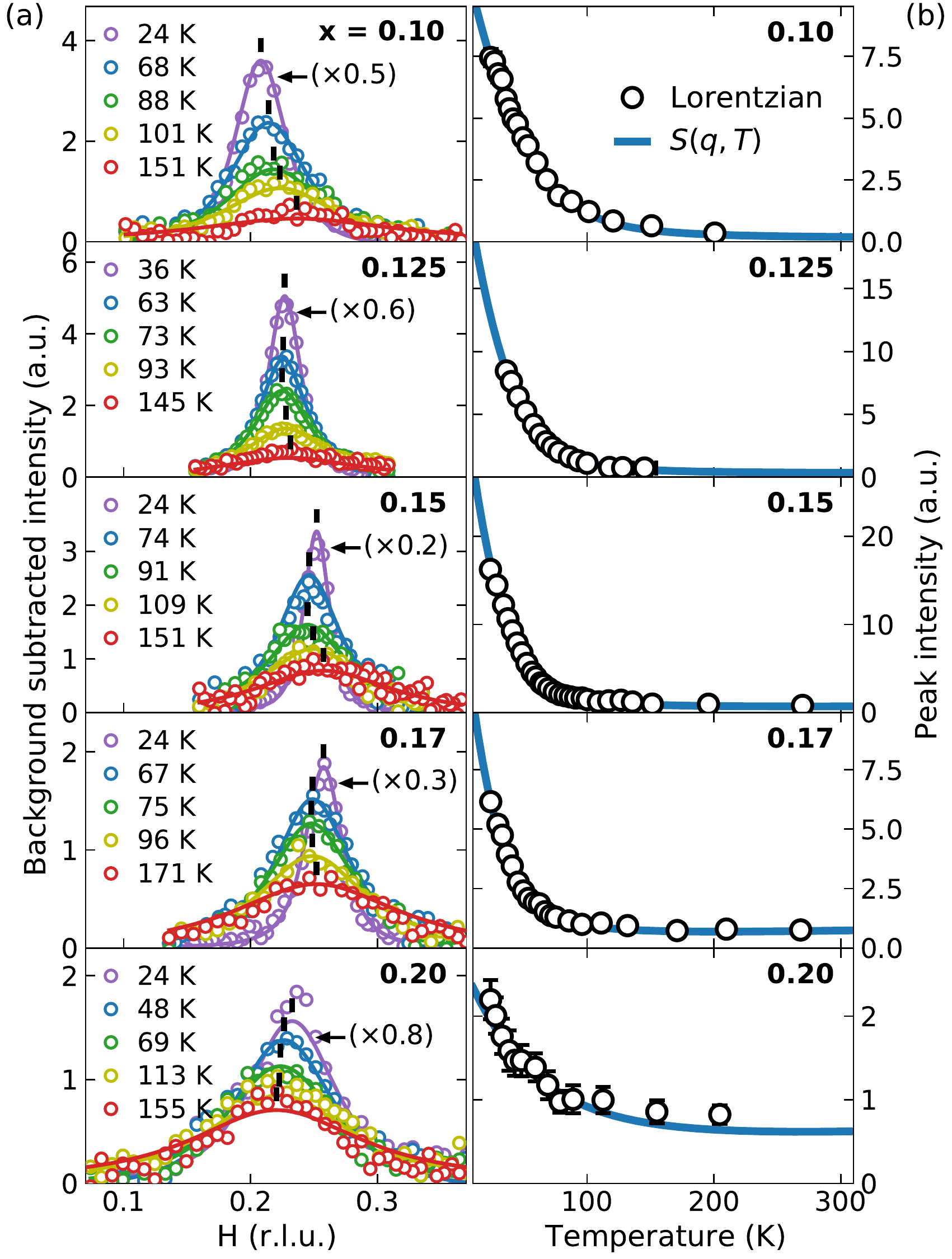}
\caption{(a) H-scans of CDW at selected temperatures with fluorescence background subtracted. Solid lines are the fitting results of $S(q, T)$ using \eqref{eq:SofQandT}. Black ticks above each curve indicate the center positions of the peaks. (b) Temperature dependence of CDW peak intensity obtained from the fits with a Lorentzian function (empty circles) and \eqref{eq:SofQandT} (solid lines).}
\label{fig:fig2}
\end{figure}

RSXS scans of the CDW in LESCO for $0.10 \leq x \leq 0.20$ at selected temperatures are shown in Fig. \ref{fig:fig2}(a). We make two main observations from the evolution of the CDW peak profiles. First, the onset of the CDW is very broad in temperature without a clear phase transition. In every sample, the CDW peak weakens and broadens on warming, but does not disappear up to the highest measured temperature, $T=270$ K in the case of $x=0.15$. Second, the CDW peak center [black ticks in Fig. \ref{fig:fig2}(a)] exhibits a peculiar dependence on doping and temperature. Upon warming, the temperature evolution of the peak center shows a monotonic increase to higher momenta at $x=0.10$, and a monotonic decrease to lower momenta at $x=0.20$. At the dopings in between, it shifts in a non-monotonic manner: it decreases upon warming from the base temperature and then increases at higher temperatures. Note that our data agree well with the high-resolution RIXS study of LESCO $x=0.125$ reported in Ref. \cite{Chang2020}, in which the elastic scattering was measured directly (see {\it SI Appendix VI}), validating that the observations described above are not artifacts arising from, for example, the inelastic background subtraction method.

\begin{figure}[hbt]
\centering
\includegraphics[width=\linewidth]{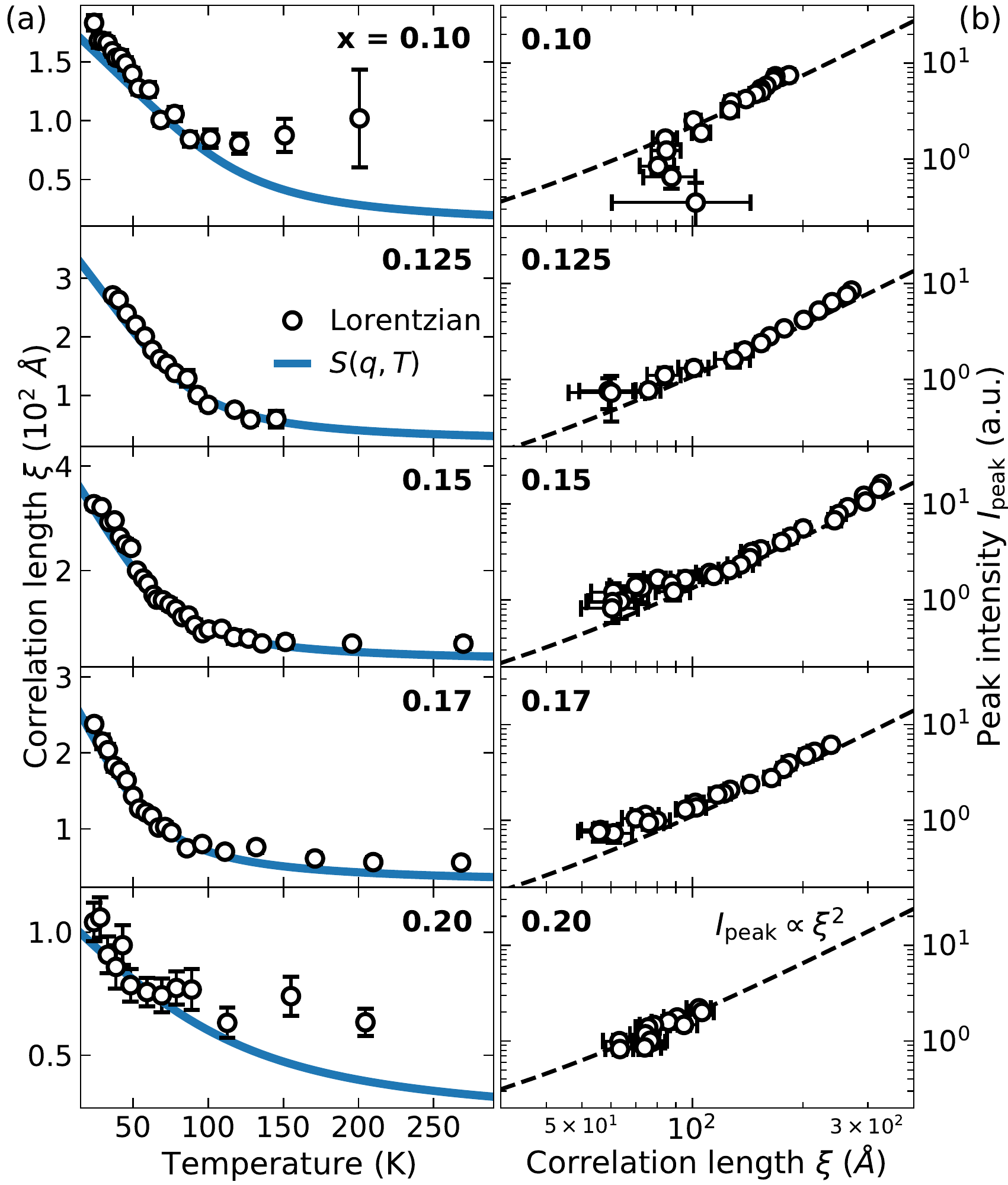}
\caption{(a) Temperature dependence of correlation length of CDW obtained from fits with a Lorentzian function (empty circles) and \eqref{eq:SofQandT} (solid lines). (b) Peak intensity plotted against correlation length in logarithmic scale. The values of peak intensity and correlation length are obtained from Lorentzian fits. Dashed lines represents the best fits of $I_{\mathrm{peak}} \propto \xi^{2}$ to the data points.}
\label{fig:fig3}
\end{figure}

\begin{figure}[hbt]
\centering
\includegraphics[width=\linewidth]{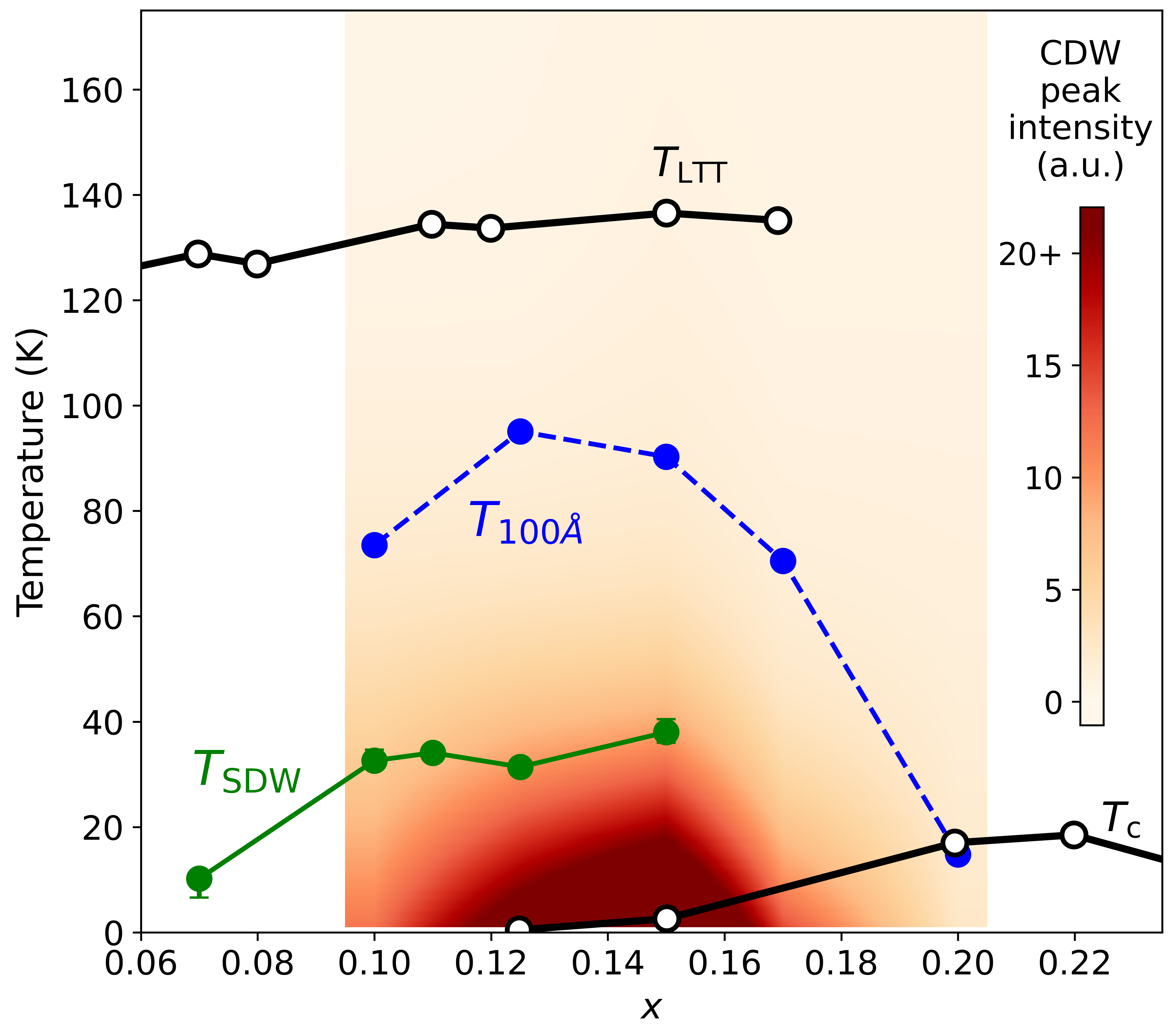}
\caption{Phase diagram of LESCO displaying the SDW onset temperature $T_{\mathrm{SDW}}$ and the CDW crossover temperature $T_{100\textrm{\AA}}$ at which the CDW correlation length takes the (arbitrary) value $\xi=100$ \AA. The overlaid color scale represents the continuous development of the CDW peak intensity $I_{\mathrm{peak}}$. The LTT structural transition temperature $T_{\textrm{LTT}}$ and superconducting transition temperature $T_{\textrm {c}}$ are taken from Ref. \cite{Fink2011}.}
\label{fig:fig4}
\end{figure}

To quantify the CDW profiles, we fit the $H$-scans at each temperature and doping with a Lorentzian function (see {\it SI Appendix V}) to obtain the peak intensity, $I_{\mathrm{peak}}$, the correlation length, $\xi$, and the $H$ component of wave vector, $Q_{\mathrm{CDW}}$. $\xi$ is defined as the ratio of the lattice parameter, $a$, to the half-width at half-maximum (HWHM) of the CDW reflection in reciprocal lattice units (r.l.u.). Figures \ref{fig:fig2}(b) and \ref{fig:fig3}(a) show the temperature dependence of $I_{\mathrm{peak}}$ and $\xi$ (black circles), illustrating a slow broadening of the CDW upon warming over a wide range of dopings. Wang et al. \cite{Chang2020} observed identical behavior at $x=0.125$ and concluded that the CDW persists far above the onset temperature reported by previous RSXS studies \cite{Fink2009, Fink2011, Achkar2016}. Our result extends this conclusion to the entire doping range, where the CDW persists above the highest temperature measured at all compositions where it is observed ($x \neq 0.07)$.

The development of the CDW reflected in the temperature dependence of $I_{\mathrm{peak}}$ and $\xi$ [Fig. \ref{fig:fig2}(b), \ref{fig:fig3}(a)] does not adhere to expectations for a traditional mean-field phase transition. For all dopings, the evolution of the spectra is gradual and we do not find any indication of a sharp or even a rounded transition. This may suggest that the CDW is weak and that the spectra indicate merely incipient CDW fluctuations. Such an interpretation is, however, at odds with the significant correlation length, approaching $\sim 100$ unit cells at low temperature (Fig. 3). Another possibility is that the nominal CDW transition temperature, in the absence of disorder, is higher than the highest temperatures measured. In this scenario, the amplitude of the CDW is already well-formed at all temperatures considered, and the gradual evolution of the spectra reflects the coupling of disorder to the CDW and the suppression of phase fluctuations as temperature is lowered.

The latter perspective is supported by our finding that $I_{\mathrm{peak}} \propto \xi^{2}$ over the range of measured temperatures [Fig. \ref{fig:fig3}(b)]. This behavior was first reported in Ref. \cite{Chang2020} for LESCO and several other La-cuprates with $x=0.125$ doping, and our results confirm this relationship over a wide range of doping from $x=0.10$ to $0.20$. The CDW in La-cuprates is known to be essentially two-dimensional, with weak out-of-plane correlations \cite{Hucker2011, Christensen2014}, and correlations in the $(H,K)$ plane that are isotropic \cite{Hucker2011, Wilkins2011}.
If a similar temperature dependence of the correlation length in the $K$ direction is assumed, this suggests that the in-plane integrated intensity, expected to be proportional to the CDW amplitude, is essentially temperature-independent.

For a quantitative test of whether phase fluctuations can account for the temperature dependence of the spectra, we consider a non-linear sigma model as described in Ref. \cite{nie-2014} and {\it SI Appendix VII}. This model assumes that the amplitude of the CDW is formed at high temperatures. In the presence of a linear coupling between disorder (parameterized by $\sigma$) and the CDW order parameter, there is no true long range CDW order below four dimensions at any temperature \cite{Imry-Ma-1975}. However, in the unbroken phase, the model predicts spectra with the lineshape
\begin{align}
S(q, T) &= T G(q, T) + \sigma^2 G^2(q, T) \label{eq:SofQandT}\\
G(q, T) &= \frac{1}{\kappa q^2 + \mu(T)}.
\end{align}
Here, $q = Q - Q_{\rm CDW}$ is momentum relative to $Q_{\rm CDW}$, $\kappa$ is the CDW stiffness, and $\mu(T)=\kappa \xi(T)^{-2}$, where $\xi(T)$ is the CDW correlation length. The temperature dependence of the lineshape is contained in $\mu(T)$, which is determined by
\begin{equation}\label{eq:muofT}
4\pi \kappa = T \ln\left[\frac{\Gamma}{\mu}\right] + \frac{\sigma^2}{\mu},
\end{equation}
which follows the self-consistency condition requiring the area under $S(q,T)$ be independent of $T$ (see {\it SI Appendix VII}). Note that this condition automatically accounts for the observed temperature independent in-plane integrated intensity. Including an overall normalization factor, the model has only 4 free parameters. We find that for every doping, without introducing temperature dependence to the parameters, the model fits well to the entire two-dimensional set of data for $S(q,T)$ using a single set of parameters (see {\it SI Appendix V}). Indeed, the solid lines in Figs. \ref{fig:fig2}(b) and \ref{fig:fig3}(a) compare remarkably well to the results of the Lorentzian fits (where parameters are temperature-dependent), except for a few points at high temperature where the Lorentzian fits are not strongly constrained. 

This analysis strongly suggests that the temperature dependence of the lineshape is predominantly related to the effects of disorder on thermal phase fluctuations of the CDW. While the CDW amplitude remains largely unchanged, at higher temperatures the correlations are short-ranged due to strong phase fluctuations. At lower temperatures, disorder pins the fluctuations and the CDW correlation length develops. 

Using the fit result of $S(q,T)$ we define a characteristic temperature $T_{100{\textrm \AA}}$ for each doping at which the correlation length takes the (arbitrary) value $\xi = 100$ \AA. 
The obtained values of $T_{100 \textrm{\AA}}$ are plotted in Fig. \ref{fig:fig4}. 
The doping evolution of $T_{100 \textrm{\AA}}$ exhibits a maximum around $x = 0.125$, similar to the previously reported CDW ``phase boundary'' in LESCO \cite{Fink2011}, as well as in other La-cuprates \cite{Hucker2011, Croft2014}, where the CDW onset temperatures were defined based on the evolution of peak intensity. 
We should emphasize that $T_{100 \textrm{\AA}}$ is not a phase boundary but a crossover scale. 
The onset of the CDW amplitude takes place at a mean-field temperature which is much higher than $T_{100\textrm{\AA}}$. At any rate, in the presence of disorder even the onset of the CDW amplitude is not a phase transition either.

\begin{figure}[hbt]
\centering
\includegraphics[width=\linewidth]{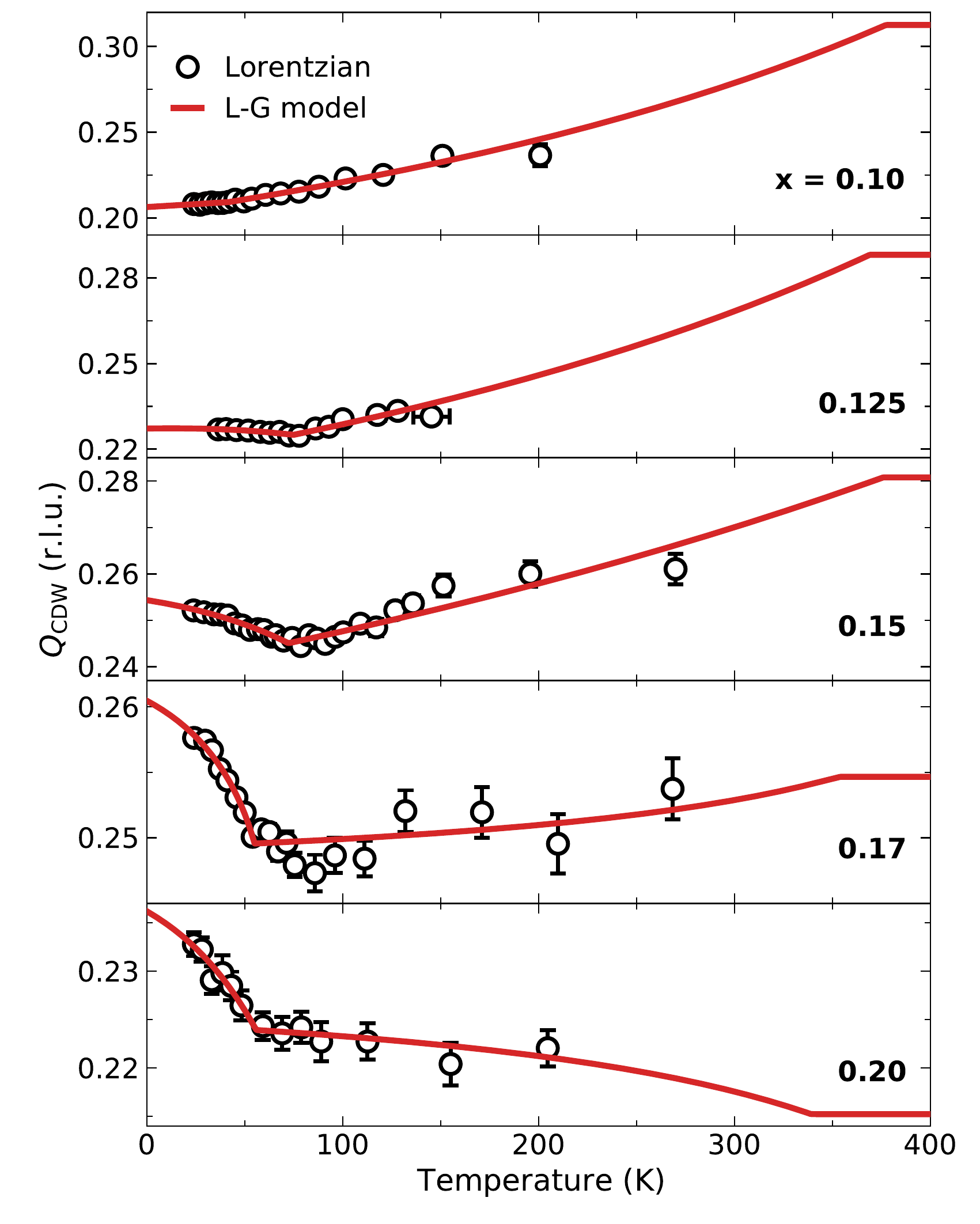}
\caption{Temperature dependence of $Q_{\mathrm{CDW}}$ obtained from Lorentzian fits (empty circles). Solid lines are fits to the Landau-Ginzburg model described in the text.}
\label{fig:fig5}
\end{figure}

\begin{figure}[hbt]
\centering
\includegraphics[width=\linewidth]{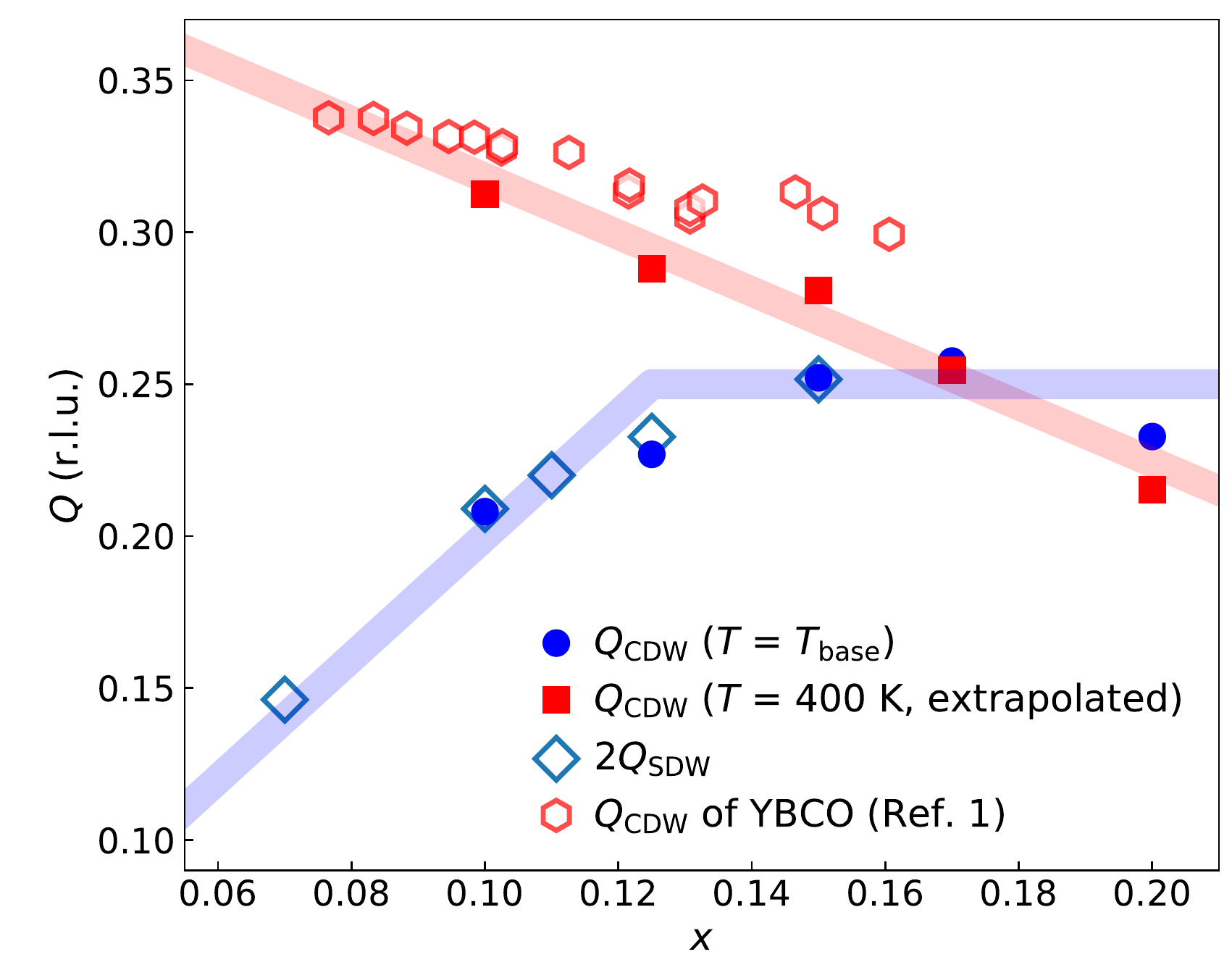}
\caption{Doping dependence of $Q_{\mathrm{CDW}}$ at the base temperature $T_\text{base}$ (blue circles) and $T$ = 400 K (red squares) predicted by the Landau-Ginzburg model described in the text. The blue solid line represents the Yamada relation and the red solid line is a guide to the eye. Twice of $Q_{\mathrm{SDW}}$ is obtained from the neutron scattering result shown in Fig. \ref{fig:fig1}(c). For a comparison, $Q_{\mathrm{CDW}}$ of {YBCO} (taken from Ref. \cite{CominReview2016}) is shown together.
}
\label{fig:fig6}
\end{figure}

We now turn to examine the non-monotonic evolution of $Q_\text{CDW}$. Unlike $I_{\mathrm{peak}}$ and $\xi$ whose temperature dependences are similar at all dopings, $Q_{\mathrm{CDW}}$ exhibits a distinct behavior for different dopings. In Fig. \ref{fig:fig5}, we display the evolution of $Q_{\mathrm{CDW}}$ as a function of $x$ and $T$ (black empty circles). At the lowest doping $x=0.10$, $Q_{\mathrm{CDW}}$ continuously increases upon warming from 0.208 r.l.u. at $T=24$ K to 0.236 r.l.u. at $T=200$ K. However at higher dopings, a kink is present around $T \sim 75$ K where the trend of the temperature dependence changes: at $x=0.15$, $Q_{\mathrm{CDW}}$ decreases from 0.251 r.l.u. at $T=24$ K to 0.244 r.l.u.  at $T=78$ K, and then increases to 0.261 r.l.u. at $T=270$ K. The increasing trend above the kink becomes less pronounced at $x=0.17$, and changes its sign at $x=0.20$, with $Q_{\mathrm{CDW}}$ decreasing over the entire temperature range from 0.233 r.l.u. at $T=24$ K to 0.222 r.l.u. at $T=207$ K, although the kink at $T \sim 60$ K is still present.

A monotonic shift of $Q_{\mathrm{CDW}}$ has been observed in several other stripe materials. In La$_{1.875}$Ba$_{0.125}$CuO$_{4}$, for example, $Q_{\mathrm{CDW}}$ moves away from its low-temperature value of 0.235 r.l.u. to a high temperature value of 0.272 r.l.u at T = 90 K \cite{Dean2017}, while in La$_{2-x}$Sr$_{x}$NiO$_{4}$ $Q_{\mathrm{CDW}}$ shifts towards a commensurate value of 1/3 upon warming regardless of doping level \cite{Tokura2001}. 
In both cases, it is believed that the coupling of the CDW to spin correlations causes $Q_{\mathrm{CDW}}$ to shift from an ``intrinsically favored'' high-temperature value with decreasing $T$. However, the non-monotonic evolution of $Q_{\mathrm{CDW}}$ found in our study is a new phenomenon, and it suggests  that there are multiple competing effects at play.

Noting that the kink in the $T$ dependence of $Q_{\rm CDW}$ occurs at temperatures comparable to the onset temperature of SDW order (Fig. 1), and that the mutual commensurability of CDW and SDW orders is known to be strong in La-based cuprates, we posit that the low-temperature behavior of $Q_{\rm CDW}$ is again related to the onset of SDW order. 
However, we attribute the behavior at intermediate $T$ to other effects. While the microscopic mechanism of CDW order in cuprates is not fully understood, we can still describe the data by an effective theory. Because of the significant variation of $Q_{\rm CDW}$ with temperature and doping, even in the absence of SDW order, such a theory must account for the finite electronic compressibility of LESCO, i.e., that it has a finite conductivity in the temperature range of interest. Such a finite compressibility is known to influence the temperature dependence of a CDW ordering wavevector \cite{Brown-2005}.  

Thus, for a minimal theoretical description of the non-monotonic behavior of $Q_{\rm CDW}$, we start with the Landau-Ginzburg theory for an incommensurate CDW in a system with finite compressibility, following the approach of Ref. \cite{Brown-2005}. We include also the free energy of the SDW order and the CDW-SDW coupling. The latter produces a tendency to mutual commensurability of the CDW and SDW ordering wavevectors. The full expression for the free energy and its implications are given in the {\it SI Appendix VII}.

The CDW ordering wavevector can be calculated by minimizing the free energy with respect to the CDW and SDW order parameters. The resulting temperature dependence of $Q_\text{CDW}$, generically, shows smooth behavior at high temperature, and a kink at the SDW transition. It is reasonable to assume the parameters may depend on $x$ though they should not change with $T$. A straight-forward multiparameter optimization results in the fit curves in Fig. \ref{fig:fig5}. These fits incorporated our knowledge from the earlier lineshape analysis (Fig. 2(a)), which showed that the integrated CDW intensity remains significant at the highest temperatures measured, by assuming a very high nominal CDW transition temperature $T_{\rm CDW} \sim 400 K$.

A striking and somewhat surprising result is obtained by using the optimized Landau-Ginzburg theory to extrapolate the value of $Q_{\rm CDW}$ outside the range of temperatures measured in the experiment. 
Figure \ref{fig:fig6} shows the values of $Q_{\mathrm{CDW}}$ at the base temperature, $T_\text{base}$, and $T$ = 400 K that are extrapolated from the fitting result using the Landau-Ginzburg model. The base temperature $T_\text{base}$ is 24 K, except for $x$ = 0.125 where it is 36 K. The doping dependence of $Q_{\mathrm{CDW}}$ at $T$ = $T_\text{base}$ roughly follows the Yamada relationship:  $Q_{\mathrm{CDW}}$ is close to twice the doping value for $x \leq \frac{1}{8}$ and saturates at 0.25 r.l.u. for $x \geq \frac{1}{8}$. This behavior can be understood as a result of the mutual commensurability between the CDW and SDW. 

The doping dependence at $T=400$ K is, however, completely different from the one at $T_\text{base}$. At $T$ = 400 K, the extrapolated value of $Q_{\text{CDW}}$ is close to 0.31 r.l.u. at $x=0.10$, and continuously {\it decreases} with doping, reaching 0.22 r.l.u. at $x=0.20$. Although the absolute values are slightly different, this trend mimics the behavior observed in YBCO and Bi-based compounds.
This observation suggests that the behavior of $Q_{\text{CDW}}$ in this La-based cuprate, if it could be measured at extremely high temperature where the effect of the SDW is minimal, would follow the same phenomenology as YBCO and Bi-based cuprates. Further, this suggests that there is a universal CDW mechanism operating in all cuprates, which simply manifests differently in different materials because of extrinsic effects such as pinning on the SDW. The key new ingredient included in our Landau-Ginzburg theory is the compressibility of the uncondensed electrons \cite{Brown-2005}, which permits a proper determination of the effect of temperature on the CDW wave vector. 

In summary, we carried out a comprehensive study of the CDW and SDW in isotopically enriched LESCO over a wide range of doping and temperature. The theoretical analysis of the RSXS data suggests that the CDW is formed at a much higher temperature than previously thought. The CDW is initially short-ranged due to thermal phase fluctuations, and a crossover takes place upon lowering temperature to the regime where disorder pins the fluctuations and the CDW correlation length develops. We also identified the new effect of charge compressibility that competes with the effect of CDW-SDW coupling in description of the non-monotonic shift of $Q_\mathrm{CDW}$ with doping and temperature. The extrapolation of the Landau-Ginzburg model reveals that the doping dependence of $Q_\mathrm{CDW}$ at $T=400$ K follows the same trend as is  observed in YBCO and Bi-based cuprates. Our study suggests that the CDW is a generic phenomenon in hole-doped cuprates with a common formation mechanism.

\subsection*{Sample growth and characterization}
Single crystals of La$_{1.8-x}$Eu$_{0.2}$Sr$_{x}$CuO$_{4}$ with $x=0.07$, 0.10, 0.11, 0.125, 0.15, 0.17, and 0.20 were grown by the traveling-solvent floating zone technique. Naturally occurring europium has a large neutron absorption cross section, which has obstructed the study of spin order in LESCO using neutron scattering. To address this issue, we have grown isotopically enriched crystals using $^{153}$Eu which has smaller neutron absorption cross section than that of the natural abundance. This isotope did not have any noticeable effect on the electronic or magnetic properties of the crystals. 

The hole doping level of the crystals were controlled by the Sr concentration, and it was confirmed by comparing the measurements of the Sr and hole concentration. The Sr concentration were measured using energy-dispersive x-ray fluorescence measurements (ED-XRF) or inductively coupled plasma optical emission spectroscopy (ICP-OES); the two techniques gave identical Sr concentration when measuring the same sample. To exclude other factors that may affect the hole doping, we also directly measured the hole concentration from the O K edge x-ray absorption spectra (XAS) \cite{CTChen1991}. The XAS were measured in total fluorescence yield mode using a transition edge sensor (TES) array detector \cite{Doriese2017, Joe2020}, which provides an excellent detection efficiency, at beamline 13-3 of the Stanford Synchrotron Radiation Lightsource (SSRL). The measured hole doping $p$ showed a good agreement with the Sr concentration $x$ (see {\it SI Appendix II}).

The Miller indices $(H, K, L)$ denote a momentum transfer wave vector $Q=(2 \pi H/a, 2 \pi K/b, 2 \pi L/c)$ in a tetragonal unit cell, where the lattice parameters are $a=b=3.79$ \AA and $c=13.14$ $\AA$ at $x=0.125$. For the full doping dependence of the lattice parameters, see {\it SI Appendix  I}.

\subsection*{RSXS measurements}
RSXS experiments at Cu $L_{3}$ edge ($\sim 932 eV$) were carried out at beamline 13-3 of the SSRL. The samples were cleaved in air to expose (0, 0, 1) surface before RSXS measurements, and ($H$, 0, $L$) plane was aligned to the scattering plane inside the scattering chamber using a 4-circle in-vacuum diffractometer. The CDW peak profiles along the in-plane momentum direction $H$ were measured by performing scans of the sample angle $\theta$ with respect to the incident x-ray beam, while the detector was fixed at 120$^\circ$. The samples were cooled using an open-cycle helium cryostat with a base temperature of 24 K. In order to subtract fluorescence backgrounds accurately, and thereby to enhance sensitivity to scatterings from CDW, we employed a two-dimensional CCD detector which measures CDW signals and backgrounds simultaneously. The background subtraction method is described in {\it SI Appendix IV} and also in Ref. \cite{Wen2019}.

\subsection*{Neutron scattering measurements}
Elastic neutron scattering measurements were performed using the HB-3 and HB-1A triple axis spectrometers at the High Flux Isotope Reactor (HFIR) at Oak Ridge National Laboratory (ORNL). For HB-3, we used an incident energy of $E = 14.7$ meV and set collimators to 48'-60'-60'-120'; for HB-1A, we used an incident energy of $E = 14.6$ meV and set collimators to $40'$-$40'$-$40'$-$80'$. We measured the samples with $x=0.07$, 0.10, 0.11, 0.125, and 0.15 on HB3, with additional data gathered on HB-1A only for samples $x=0.10$, 0.125, and 0.15. These crystals were mounted in the ($H$, $K$, 0) scattering plane inside a helium cryostat with a base temperature of 1.5 K; we used a closed cycle refrigerator with a base temperature of 4 K for the $x=0.11$ sample. The data from HB-1A were scaled by an overall constant multiplication factor to match the data from HB3, and the two data sets agreed well with each other.


\begin{acknowledgements}
X-ray experiments were supported
by the U.S. Department of Energy, Office of Basic Energy Sciences grant no. DE-FG02-06ER46285.
Use of the SSRL was supported by DOE contract DE-AC02-76SF00515.
Neutron scattering experiments and growth of LESCO crystals were supported by DOE BES grant no. DE-SC0012368. The neutron measurements used the High Flux Isotope Reactor, a DOE Office of Science User Facility operated by Oak Ridge National Laboratory. 
Theoretical work was supported by National Science Foundation grant DMR-1725401 (EF) and DOE BES grant no. DEAC02-76SF00515 (SAK).  
We acknowledge support from the  Gordon and Betty Moore
Foundation's EPiQS Initiative through grants GBMF9452 (PA), GBMF4305 (EWH) and GMBF8691 (EWH).
\end{acknowledgements}



%

\onecolumngrid

\appendix
\section{Supporting Information Appendix (SI)}

\section{I. Lattice parameters}
Table \ref{tab:Latt} shows lattice parameters of the La$_{1.8-x}$Eu$_{0.2}$Sr$_{x}$CuO$_{4}$ (LESCO) samples with $x=0.07$, 0.10, 0.125, 0.15, 0.17, and 0.20 with a tetragonal unit cell ($a=b$, $\alpha=\beta=\gamma=90^{\circ}$). The lattice parameters are determined from single crystal x-ray diffraction measurements with Cu K$_{\alpha}$ x-ray source. 

\begin{table}[hbt]
\caption{\label{tab:Latt} Lattice parameters of LESCO with $0.07 \leq x \leq 0.20$.}
\begin{tabular}{|c|ll|} 
\hline
\hspace{1cm}$x$\hspace{1cm} & \hspace{1cm}a (\AA) & \hspace{1cm}c (\AA)\hspace{1cm} \\
\hline
0.07 & \hspace{1cm}3.806(2) & \hspace{1cm}13.139(1)\hspace{1cm} \\
0.10 & \hspace{1cm}3.79(1) & \hspace{1cm}13.137(3)\hspace{1cm} \\
0.125 & \hspace{1cm}3.79(1) & \hspace{1cm}13.136(4)\hspace{1cm} \\
0.15 & \hspace{1cm}3.79(1) & \hspace{1cm}13.161(5)\hspace{1cm} \\
0.17 & \hspace{1cm}3.79(1) & \hspace{1cm}13.166(4)\hspace{1cm} \\
0.20 & \hspace{1cm}3.77(1) & \hspace{1cm}13.181(3)\hspace{1cm} \\
\hline
\end{tabular}
\end{table}

\section{II. Measurement of hole doping $p$}
\begin{figure}[hbt]
\includegraphics[width=0.5\textwidth]{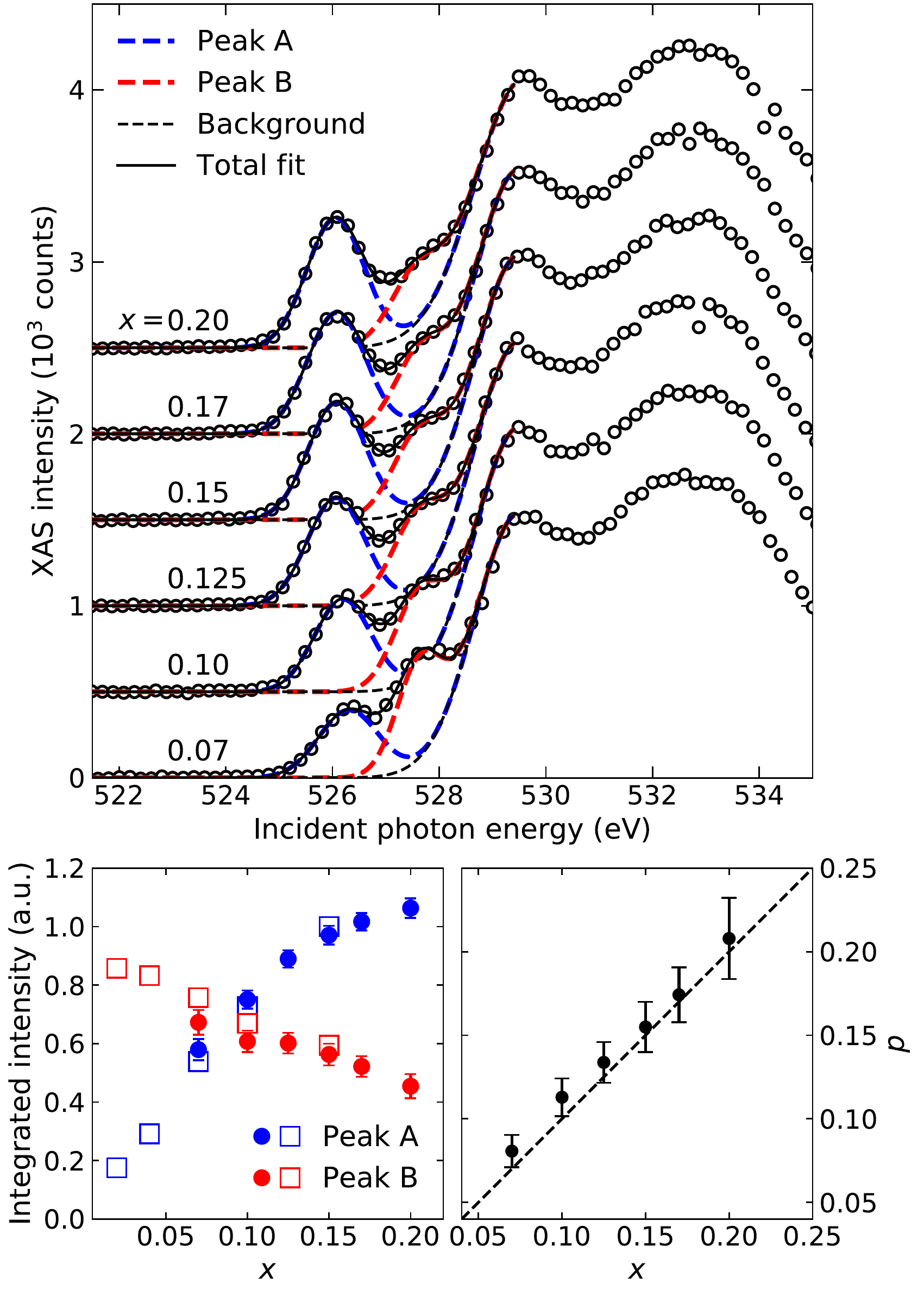}
\caption{\label{fig:XAS}
(a) Oxygen K-edge XAS spectra of LESCO samples. The pre-edge regions of the spectra exhibit two distinct peaks which are fitted with two Gaussian functions (peak A: blue dashed curves, peak B: red dashed curves) and a background (gray dashed curves). The total lineshapes (solid lines) fit well to the XAS data points. (b) Doping evolution of integrated intensities of peak A and B (filled circles). For a comparison, data points from the same analysis on LSCO in Ref. \cite{CTChen1991} are plotted together (hollow squares). (c) Measured hole doping $p$ plotted against Sr concentration $x$. Dashed line represents $p=x$.
}
\end{figure}

Hole doping levels of the LESCO samples are mainly determined by two factors: Sr concentration and oxygen non-stoichiometry. In order to confirm that the doping is controlled by Sr concentration and the effect of oxygen non-stoichiometry is minimal, we have performed x-ray absortpion spectroscopy (XAS) experiments as a direct measurement of hole concentration of the samples. It has been observed that O K edge XAS spectra of hole-doped high-T$_{c}$ cuprates exhibit two pre-edge peaks and the intensity of the lower energy peak increases linearly with hole concentration in the doping range of 0 to 20\% \cite{CTChen1991, Pellegrin1993, Peets2009, CCChen2013}. Figure \ref{fig:XAS}(a) shows the O K edge XAS spectra of the LESCO samples that were used in this study, which resemble the spectra of La$_{2-x}$Sr$_{x}$CuO$_{4}$ reported in Ref. \cite{CTChen1991}. The pre-edge regions of the result are analyzed in the same way described in Ref. \cite{CTChen1991}, where the background is modeled with a Gaussian function and the two pre-edge peaks (the lower-energy peak: peak A, the higher-energy peak: peak B) are fitted with two Gaussian lineshapes, and the integrated intensities of peak A and B are obtained [Fig. \ref{fig:XAS}(b)]. According to Ref. \cite{CTChen1991}, the relationship between the ratio of the intensities of two pre-edge peaks, $I_{\mathrm{peak A}}/I_{\mathrm{peak B}}$ and the hole doping $p$ is found to be $I_{\mathrm{peak A}}/I_{\mathrm{peak B}} = (11.6 \pm 0.4)p-(0.076 \pm 0.03)$. Using this relationship, we estimated the doping level $p$ of the LESCO samples [Fig. \ref{fig:XAS}(c)] and it agrees well with Sr concentration $x$.

\section{III. SDW $H$-scan and onset temperature}
\begin{figure*}[hbt]
\includegraphics[width=0.6\textwidth]{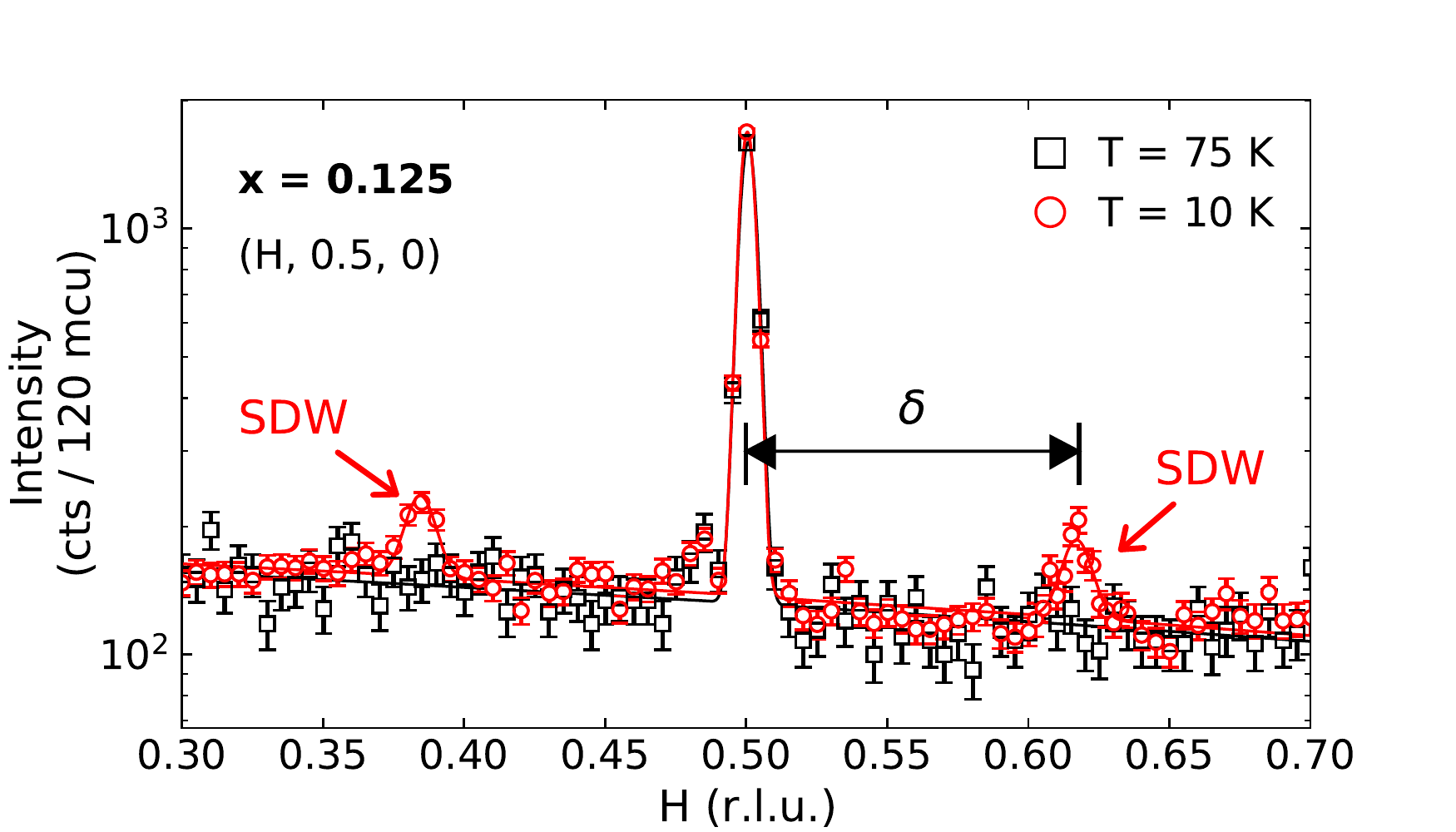}
\caption{\label{fig:SDWwidescan}
$H$-scans of SDW of LESCO $x=0.125$ at temperatures 10 K and 75 K. The SDW peaks are the satellite peaks. The (0.5, 0.5, 0) peak we attribute to multiple scattering. the solid lines are fits to Gaussian functions. 
}
\end{figure*}

\begin{figure*}[hbt]
\includegraphics[width=5.7in]{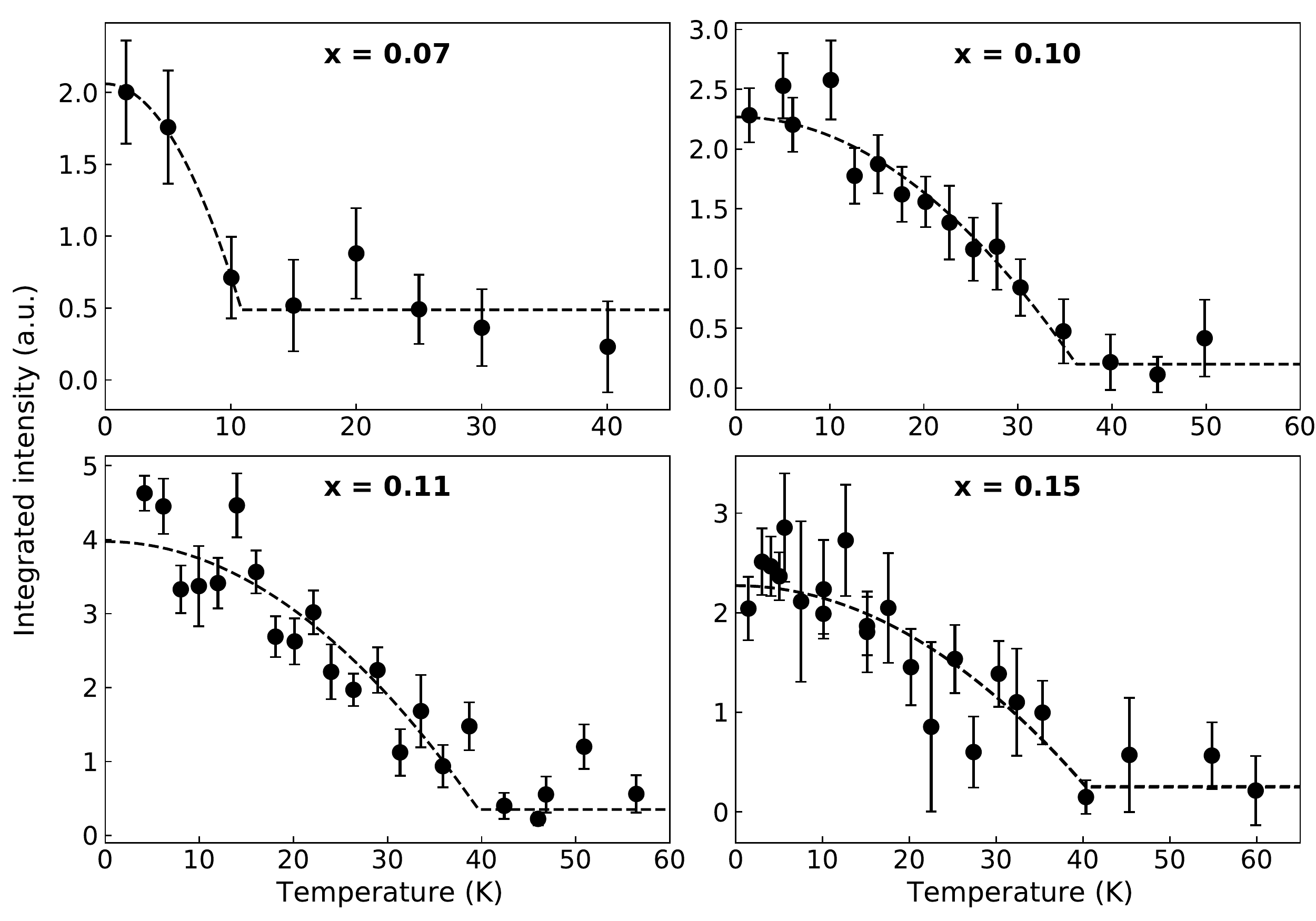}
\caption{\label{fig:SDWvsT}
Integrated intensity of SDW of LESCO $x=0.07$, 0.10, 0.11, and 0.15. The dashed line shows a fit to Eq. [1] in the main manuscript. Error bars correspond to standard errors. 
}
\end{figure*}

Figure \ref{fig:SDWwidescan} shows the $H$-scan of SDW in LESCO $x=0.125$, where the satellite SDW peaks are present on both side of (0.5, 0.5, 0). The SDW peaks were seen in equivalent points in other Brillouin zones, too. In Fig. \ref{fig:SDWvsT}, the temperature dependence of the integrated intensity of SDW is presented for $x=0.07$, 0.10, 0.11, and 0.15. The integrated intensities are fitted with the square-root model described in the main manuscript to infer the onset temperatures. The data for $x=0.125$ is presented in the main manuscript.

\section{IV. RSXS Background subtraction}
\begin{figure*}[hbt]
\includegraphics{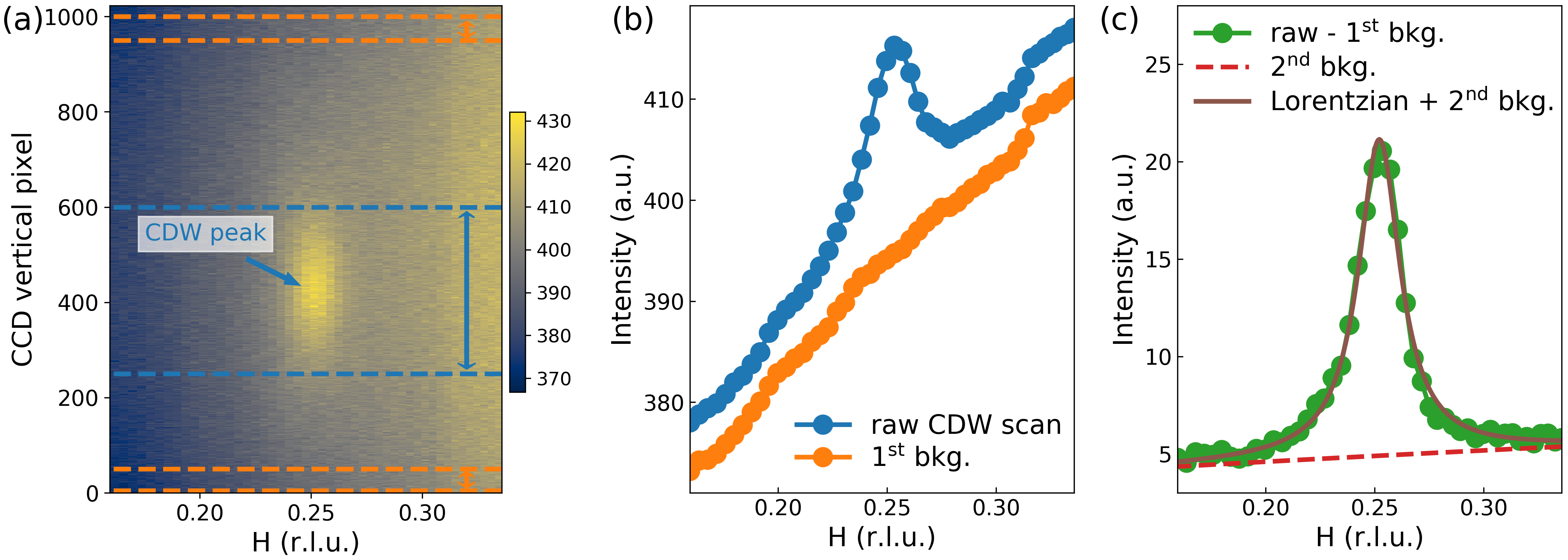}
\caption{\label{fig:bkg}
(a) Intensity map of $H$-scan of CDW of LESCO $x$=0.15 at $T$ = 23 K measured using a CCD detector. The region within the blue (orange) dashed lines captures CDW (fluorescence) signals. (b) Raw CDW $H$-scan and the first background curve obtained from the blue and orange region, respectively, in (a). (c) The result of the first background subtraction (green curve) that is fitted with the second background modeled with a quadratic function (red dashed line) and a Lorentzian function (brown curve).
}
\end{figure*}

Fluorescence background subtraction of RSXS measurements is performed utilizing a large area CCD detector that measures both CDW and fluorescence signals simultaneously, which greatly enhanced the detection sensitivity \cite{Wen2019}. The subtraction is done in two steps. As shown in Fig. \ref{fig:bkg}(a), the center part of the CCD images (the region within blue dashed lines) captures the CDW peak profile, while the top and bottom parts (the regions with in orange dashed lines) are far away from the CDW peak and essentially capturing the $H$ dependence of fluorescence intensity. Figure \ref{fig:bkg}(b) shows the raw $H$-scan of CDW (blue curve) and the fluorescence background (orange curve) obtained by averaging intensities within each region of the images, and the first background subtraction is carried out by subtracting the two curves. In Fig. \ref{fig:bkg}(c), the result of the first background subtraction (green curve) is plotted, which shows a residual background due to the vertical variation of fluorescence in the CCD images. The remaining background is modeled with a quadratic function (red dashed line) and used for the second background subtraction. The final result after the first and second background subtraction is modeled with a Lorentzian function (brown curve) which fits well to the data.

\section{V. Lorentzian fits and $S(q, T)$ fits to RSXS scans}
\begin{figure*}[tbhp]
\includegraphics[width=\linewidth]{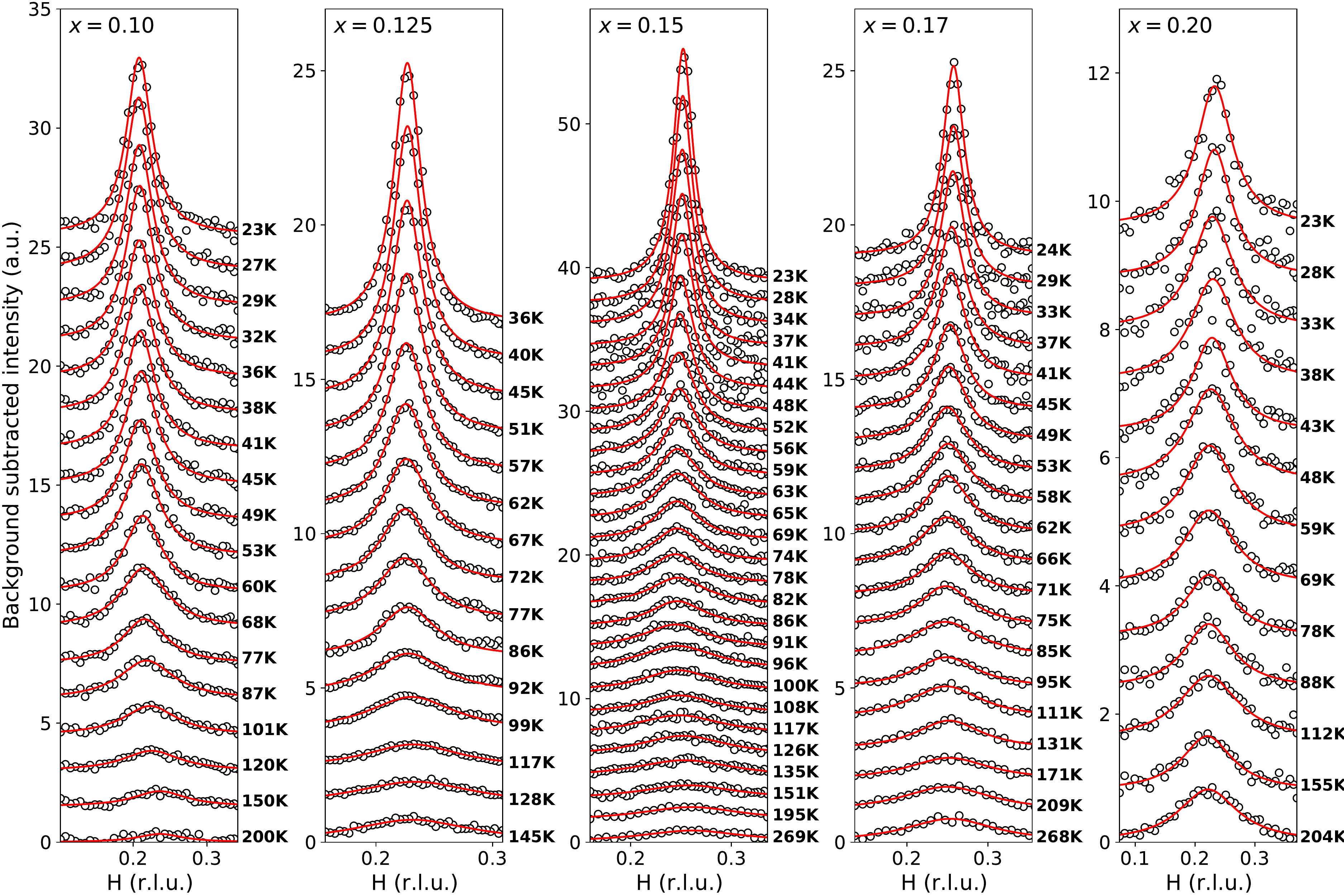}
\caption{\label{fig:LESCOscanslorentzian}
Background subtracted $H$-scans of CDW of LESCO $x$=0.10, 0.125, 0.15, 0.17 and 0.20 at a wide range of temperature. The solid lines are Lorentzian fits.
}
\end{figure*}

\begin{figure*}[hbt]
\includegraphics[width=\linewidth]{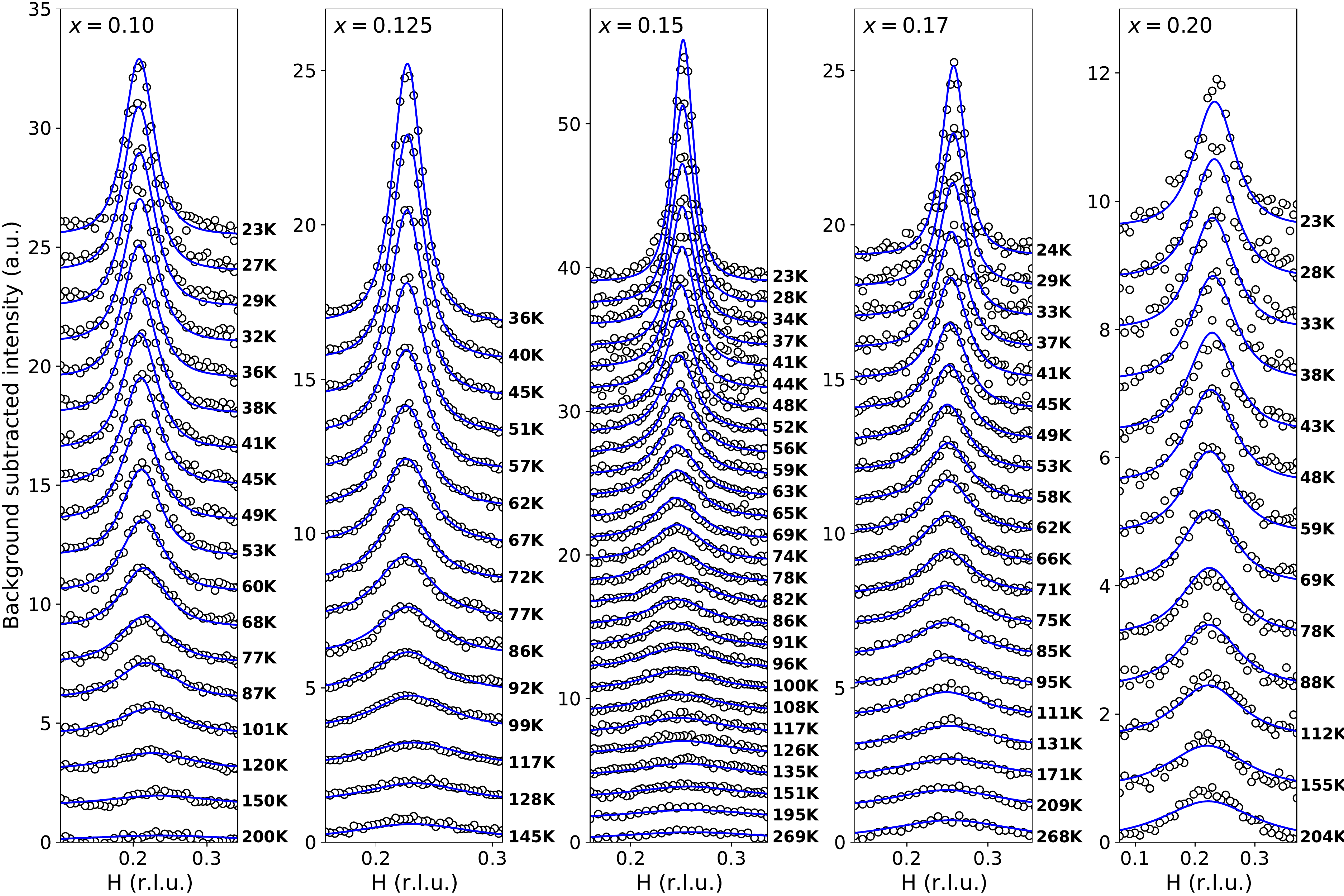}
\caption{\label{fig:LESCOscanssofq}
Background subtracted $H$-scans of CDW of LESCO $x$=0.10, 0.125, 0.15, 0.17 and 0.20 at a wide range of temperature. The solid lines are fits to $S(q,T)$ given in Eq (1) in the main manuscript.
}
\end{figure*}

Figure \ref{fig:LESCOscanslorentzian} shows the $H$-scans of CDW after background subtraction and Lorentzian fits from which we obtained the peak intensity, correlation length, and wave vector of CDW at each temperature and doping. Figure \ref{fig:LESCOscanssofq} shows the fitting result with $S(q,T)$ given in Eq. (1) in the main manuscript. The quality of the fit is very good, indicating that the model describes the entire two-dimensional set of $H$-scans (as a function of $q$ and $T$) very well using a single set of temperature-independent parameters.
 
\section{VI. Comparison to a RIXS result}
\begin{figure}[tbhp]
\includegraphics[width=0.6\linewidth]{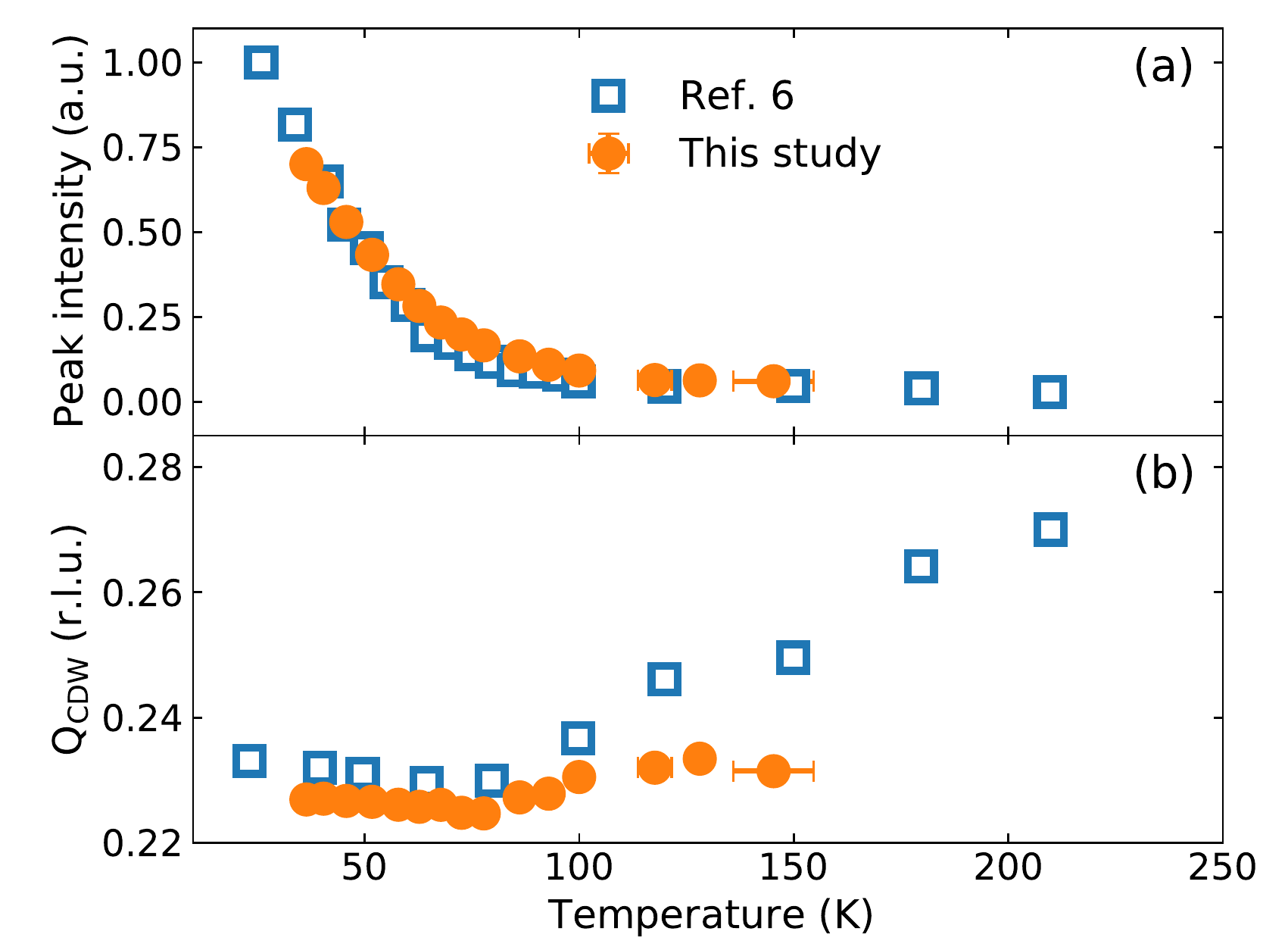}
\caption{\label{fig:RIXS}
Comparison of temperature dependence of (a) CDW peak intensity and (b) $Q_{\mathrm{CDW}}$ in LESCO $x$=0.125 reported in Ref. \cite{Chang2020} (hollow squares) and measured in this study (filled circles).
}
\end{figure}

In order to cross-check the results of this study, the temperature dependence of CDW peak intensity and $Q_{\mathrm{CDW}}$ are compared with a previously reported result. Figure \ref{fig:RIXS} shows a comparison with the data from Ref. \cite{Chang2020} where a high-resolution ($\Delta E \geq 19$ meV) resonant inelastic x-ray scattering measurement on LESCO $x$=0.125 is reported. The data from this study and Ref. \cite{Chang2020} agrees well with each other. Especially, both results exhibit a gradual increase of CDW peak intensity and a kink at $T \sim 75$ K in the temperature evolution of $Q_{\mathrm{CDW}}$. 

\clearpage
\section{VII. Supplementary Theory Material}
\subsection{Summary of experimental facts}

The experimental results from x-ray diffraction and  neutron scattering are summarized as follows:
\begin{enumerate}
\item
X-ray diffraction studies of Eu-doped {\LSCO} (LESCO) samples with doping in the range $0.1 < x <0.2$ and temperatures in the range $15 K < T < 250 K$ show CDW order.
\item
The CDW order is unidirectional with an ordering wave vector ${\bm Q}_{\rm CDW}=Q_{\rm CDW} {\hat {\bm a}}$ that varies with doping and with temperature. Here ${\hat {\bm a}}$ is a unit vector along the $a$ direction of the Cu-O plane. 
\item
The onset of CDW order is broad, and essentially the same for all dopings. The CDW order  it is well developed well above the structural transition and it is essentially unaffected by it.
\item
For much of the measured temperature range the integrated intensity of the CDW peak is essentially temperature-independent, indicating that the strength of the CDW order parameter (i.e. its amplitude) is essentially constant in the measured range of temperatures (although it is different at different doping levels).
\item
Neutron scattering data shows that these samples exhibit SDW order below a temperature which depends on doping. 
\item
The SDW transition is fairly broad, the order is unidirectional with an ordering wave vector $Q_{SDW}$ that appears to be temperature-independent and doping-dependent. 
\item
Except for the very underdoped sample (with $x\sim 0.1$) the temperature dependence of the CDW ordering wave vector $Q_{\rm CDW}$ is not monotonic, decreasing with lowering temperatures down to minimum below which it increases rapidly. 
\item
The minimum of $Q_{\rm CDW}$ appears to occur near the onset of the SDW.
\item
At low temperatures the (incommensurate) CDW and SDW orders become  commensurate with each other and exhibit the characteristic relation $Q_{\rm CDW} = 2 Q_{\rm SDW}$ observed in {\LBCO} (LBCO) and in {\LNSCO} (LNSCO).
\end{enumerate}

Several conclusions emerge from these experimental results.
The fact that, for given doping level,  the amplitude of the order parameter is essentially independent of the temperature suggests that the thermal fluctuations of the {\em phase} of the CDW order parameter play a dominant role (at least for some of the phenomena), i.e. a non-linear sigma model approach may give a good description. Also,
the fact that the CDW thermal transition is broad implies that disorder plays a key role as a ``random field'' effect coupled linearly to the CDW order parameter. This approach is presented below.
On the other hand, the fact that the CDW ordering wave vector varies significantly with temperature and doping strongly suggests that a finite electronic compressibility must be part of the explanation of the observed phenomena. This effect is not easily accounted for in a simple non-linear sigma model approach.
Similarly, the theory must also describe the mutual commensurability of the CDW and the SDW orders, which is more easily described in a landau-Ginzburg approach, as is the temperature dependence of $Q_{\rm CDW}$. For the sake of simplicity we will use both approaches.

\subsubsection{Landau-Ginzburg Theory}

\subsubsection{CDW alone}

 We will consider first the case when there is only CDW order. The Landau-Ginzburg theory for a CDW in a system with finite charge compressibility was discussed in Ref. [\cite{Brown-2005}], which is an extension of the standard Landau theory for CDW due to McMillan [\cite{mcmillan-1975}] for a system with finite compressibility. We will consider a 3D system with unidirectional order, with ordering wave vector ${\bm Q}=Q  {\hat {\bm Q}}$. Here follow closely Ref.[\cite{Brown-2005}].
 
 In a  CDW phase the local charge density has the expansion
\begin{equation}
\rho({\bm x})=\psi_0({\bm x})+i \Lambda \psi^*({\bm x}) {\hat {\bm Q}} \cdot {\bm \nabla} \psi({\bm x})+ {\rm c.c.}
+\psi({\bm x}) \exp(i {\bm Q} \cdot {\bm x})+\psi^*({\bm x}) \exp(-i {\bm Q} \cdot {\bm x})+ \ldots
\label{eq:rho}
\end{equation}
The complex field $\psi({\bm x})$ is the order parameter of the CDW and we have kept only the leading harmonic. The real field $\psi_0(\bm x)$ represents the long wavelength fluctuations of the charge density and represents the finite  charge compressibility. The second term in Eq.\eqref{eq:rho} is McMillan's ``CDW compression term'' and describes the contribution of the local fluctuations of the CDW order parameter to the long wavelength charge fluctuations. 
Finally, ${\bm Q}$ is the ``ideal'' CDW ordering wave vector. 

We will see below that the finite compressibility leads to temperature-dependent deviations of the actual ordering wave vector from the ideal value (McMillan's theory this effect is introduced ``by hand'').
The average charge density is given by $L^3 {\bar \rho}=\int d^3x \rho({\bm x})$. We will assume that the charge density has a preferred ``normal'' density $\rho_N$. The difference $\Delta {\bar \rho}={\bar \rho}-\rho_N$ will play the role of a control parameter and, qualitatively, can be thought of as representing the doping level.

The CDW Landau-Ginzburg free energy is
\begin{align}
F_{\rm CDW}=& \int d^3x \Big\{ \frac{1}{2} K_{c} \Big| \Big(i {\hat {\bm Q}} \cdot {\bm \nabla}+\delta_0\Big) \psi({\bm x})\Big|^2
+\frac{r_c}{2} |\psi({\bm x})|^2+u_c |\psi({\bm x})|^4  \Big\}
+ \int d^3x \frac{\kappa_0}{2} (\psi_0({\bm x})-\rho_N)^2\nonumber\\
+&\frac{1}{2} \int d^3x \int d^3y (\rho({\bm x})-{\bar \rho}) V(|{\bm x}-{\bm y}|) (\rho({\bm y})-{\bar \rho})
\label{eq:LG-CDW}
\end{align}
Here $K_{c}$ is the stiffness of the CDW order parameter, $\delta_0$ is the ``pressure'', $r_c=T-T_{\rm CDW}^0$ (with $T_{\rm CDW}^0$ being the usual mean field CDW critical temperature), and $u_c$ are the usual parameters of the Landau potential.

In the free energy of Eq.\eqref{eq:LG-CDW}  $\kappa_0$ is the uniform charge compressibility and plays an important role in the physics: for $\kappa_0 \to \infty$ there is no screening and for $\kappa_0 \to 0$ the interaction becomes ultra-local. The last term represents the long-range Coulomb interaction of the charge configurations $\rho({\bm x})$ given in Eq.\eqref{eq:rho}, and
\begin{equation}
V(|{\bm x}-{\bm y}|)=\frac{e^2}{\varepsilon} \frac{1}{|{\bm x}-{\bm y}|}
\end{equation}
where $\varepsilon$ is the dielectric constant. 

The equilibrium state is obtained by the extrema of the free energy of Eq.\eqref{eq:LG-CDW} with respect to the uniform component $\psi_0$ and with respect to the CDW order parameter $\psi$. Among other things, the fluctuations of the uniform component lead to the screening of the long-range Coulomb interaction, with a screening length $\xi=\left(\frac{4\pi \varepsilon}{e^2} \kappa_0 \right)^{1/2}$ tuned by the compressibility $\kappa_0$. Thus, if $\kappa_0 \to \infty$ there is no screening and if $\kappa_0 \to 0$ the interaction is ultra-local.

We will look for the extrema of the LG free energy with an ICDW ansatz
$\psi(x)=\phi_0  \exp(i {\hat {\bm Q}}\cdot {\bm x} \delta)$, where $\delta$ is the incommensuration and $\phi_0$ is the CDW amplitude. 
Hence, the actual CDW ordering wave vector has a magnitude $Q_{\rm CDW}=Q+\delta$, along the direction ${\hat {\bm Q}}$. 

We will need to determine also the average value of the compressible charge density ${\bar \psi}_0$. In this state ${\bar \psi}_0$ is given by
\begin{equation}
{\bar \psi}_0={\bar \rho}-2\Lambda |\phi_0|^2 \delta
\label{eq:bar-psi0}
\end{equation}
which shows that the incommensuration $\delta$ shifts the average value of the compressible charge density away from the average density $\bar \rho$.

In the normal state $\psi=0$, $\rho({\bm x})={\bar \rho}={\bar \psi}_0$. The free energy in the normal state is $F_N=\frac{1}{2} L^3 \kappa_0(\Delta {\bar \rho})^2$.

The free energy density in the CDW state becomes
\begin{equation}
f_{\rm CDW}=\frac{1}{2} K_{c} (\delta-\delta_0)^2 |\phi_0|^2+\frac{1}{2} \left(r_c+2 {\tilde V}_{\rm eff}(Q_{\rm CDW})\right) |\phi_0|^2+u_c |\phi_0|^4
+\frac{\kappa_0}{2} (\Delta {\bar \rho}+2\Lambda |\phi_0|^2 \delta)^2
\label{eq:f-ICDW}
\end{equation}
where 
\begin{equation}
{\tilde V}_{\rm eff}(Q_{\rm CDW})=\frac{\frac{e^2}{4\pi \varepsilon} }{Q_{\rm CDW}^2+\xi^{-2}}
\end{equation}
 is the Fourier transform of the screen Coulomb interaction at the wave vector $Q_{\rm CDW}$ and $\xi$ is the screening length. 

Upon extremizing with respect to $\delta$ the free energy density in the ICDW state, Eq.\eqref{eq:f-ICDW}, we obtain
\begin{equation}
\delta=\delta_0 \left(\frac{b^2-a^2}{b^2+|\phi_0|^2}\right)
\label{eq:delta}
\end{equation}
where we defined
\begin{equation}
a^2=\frac{\Delta {\bar \rho}}{2\Lambda \delta_0}, \qquad b^2=\frac{K_{c}}{4\Lambda^2 \kappa_0}
\label{eq:ab}
\end{equation}
Eq.\eqref{eq:delta}  shows that the incommensuration $\delta$ depends explicitly on $\phi_0$,  the amplitude of the CDW, which depends on the temperature.  Thus, as the amplitude of the CDW order parameter $\phi_0$ increases (with decreasing temperature), the value of the incommensuration $\delta$ will decrease (increase) for $b>a$ (for $b<a$).  Notice that in the ultra-local limit, where $\kappa_0 \to 0$, $\delta \to \delta_0$ and has no temperature dependence.

By plugging in the expression for $\delta$ of Eq.\eqref{eq:delta} back into of Eq.\eqref{eq:f-ICDW}, we obtain the expression for the free energy density [\cite{Brown-2005}],
\begin{equation}
f_{\rm CDW}=\frac{1}{2} K_{c} \delta_0^2 \; \frac{(a^2+|\phi_0|^2)^2}{b^2+|\phi_0|^2}+\frac{1}{2} (r_c+2{\tilde V}_{\rm eff}(Q_{\rm CDW})) |\phi_0|^2+u_c |\phi_0|^4
\label{eq:f-ICDW-2}
\end{equation}
This expression differs from the conventional expression for the CDW free energy density in two respects. 
A minor one is the shift of the coefficient $r_c=T-T^0_{\rm CDW}$ by the Coulomb potential at the CDW wave vector. 
This shift means that the naive CDW critical temperature is suppressed by the Coulomb effect. 
However, in addition we now have the much more non-linear first term which plays a key role. 
The origin of this term is the finite compressibility. In fact, as $\kappa_0$ increases, where screening becomes poorer,
 this term shifts the critical temperature of the CDW to high temperature.
 
 \begin{figure}[hbt]
 \begin{center}
 \includegraphics[width=0.6\textwidth]{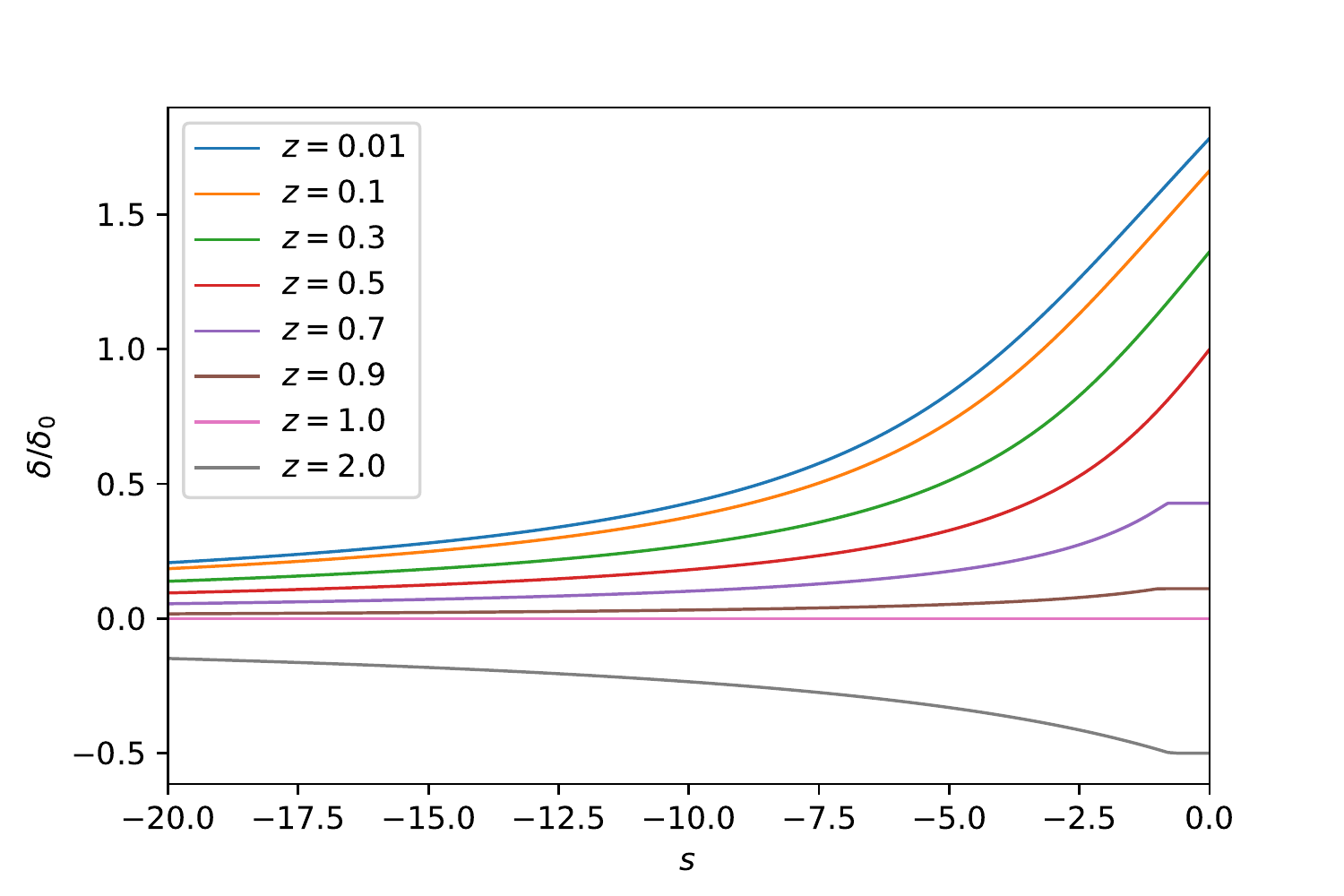}
 \end{center}
 \caption{Incommensurability $\delta$ as a function of temperature (represented by $s$) for different values of $z$, controlled by the compressibility $\kappa_0$.}
 \label{fig:Q-cdw-only}
 \end{figure}

 A numerical minimization of the CDW free energy density, Eq.\eqref{eq:f-ICDW-2}, yields a plot of the CDW amplitude $\phi_0$ and of the incommensurability $\delta$ as a function of temperature (shown in Fig.\ref{fig:Q-cdw-only}). In the figure a plot of the data for the discommensuration $\frac{\delta}{\delta_0}$, Eq.\eqref{eq:delta} is shown parametrized by the dimensionless quantities $z=\frac{b^2}{a^2}$ (with $a$ and $b$ defined in Eq.\eqref{eq:ab}) and $x=\frac{|\phi_0|^2}{b^2}$, and $s=(r_c+2{\tilde V}_{\rm eff}(Q_{\rm CDW})/K_{c} \delta_0^2$.

The upshot of these results is that a monotonic temperature-dependence of the CDW ordering wave vector can be explained as a consequence of the finite compressibility. In addition, the extrapolated value of the ordering wave vector at the nominal onset of CDW order can be related to the doping level, qualitatively represented  by $\Delta {\bar \rho}$, and hence by $a^2$. However, these expressions alone cannot describe the non-monotonic behavior seen in the experiments. Next we will see that the non-monotonicity can be explained by the interaction between the CDW and the SDW.

\subsubsection{CDW+SDW}

We now turn to the full problem, including the SDW and its interaction with the CDW. In an SDW state local magnetization ${\bm M}({\bm x})$ has an expansion in harmonics of the ordering wave vector ${\bm K}$ of the form
\begin{equation}
{\bm M}({\bm x})={\bm S}({\bm x}) \exp(i{\bm K} \cdot {\bm x})+{\rm c.c.}
\end{equation}
where we kept only the leading harmonic. The SDW order parameter ${\bm S}({\bm x})$ is a three-component complex field. In its simplest form the free energy of the SDW order parameter is (see Refs. [\cite{zachar-1998}] and [\cite{nie-2017}])
\begin{equation}
F_{\rm SDW}=\int d^3x \Big\{ \frac{1}{2} K_s \left| {\bm \nabla} S_a({\bm x})\right|^2+\frac{1}{2} r_s |{\bm S}({\bm x})|^2+u_s |{\bm S}({\bm x})|^4+{\tilde u}_s |{\bm S}^*({\bm x}) \times {\bm S}({\bm x})|^2\Big\}
\label{eq:LG-SDW}
\end{equation}
(with $a=1,2,3$) where $r_s=T-T_0^{\rm SDW}$ and $u_s>0$. We will be interested only in collinear spin order and we will set ${\tilde u}_s=0$.

The free energy of the full system, involving both charge and spin orders, is
\begin{equation}
F=F_{\rm CDW}+F_{\rm SDW}+F_{\rm int}
\label{eq:F-full}
\end{equation}
where $F_{\rm CDW}$ is given in Eq.\eqref{eq:LG-CDW} and $F_{\rm SDW}$ in Eq.\eqref{eq:LG-SDW}. The third term in Eq.\eqref{eq:F-full} is
\begin{equation}
F_{\rm int}=\int d^3x \Big\{\lambda \psi^*({\bm x}) {\bm S}({\bm x}) \cdot {\bm S}({\bm x}) \exp(i (2 {\bm K}-{\bm Q}) \cdot {\bm x})+{\rm c.c.}+\gamma |\psi({\bm x})|^2 |{\bm S}({\bm x})|^2 \Big\}
\label{eq:LG-int}
\end{equation}
Here, $\lambda$ and $\gamma$ are two coupling constants. The important coupling is $\lambda$ which describes the effects of mutual commensurability of the CDW and SDW orders. We will only consider the case in which the SDW onsets at temperatures well below the CDW, and so we will take $T_{\rm SDW}^0 \ll T^0_{\rm CDW}$, which is the experimentally relevant situation in {LESCO}. Thus, the CDW order is well developed at the temperature for the onset of the SDW.

It is convenient to parametrize the magnitudes of the ordering wave vectors of the coupled CDW and SDW as follows
\begin{equation}
{Q}_{\rm CDW}={Q}+2 {\ell}, \qquad {K}_{\rm SDW}={K}+{\ell}-{q}
\label{eq:order_wavevec}
\end{equation}
where we used the notation $\delta=2\ell$. Here $q$ measures  the mutual incommensurability of the CDW and the SDW.

The mean field ansatz now is
\begin{equation}
\psi({\bm x})=\phi_0 \exp(i 2 \ell \; {\hat {\bm Q}} \cdot {\bm x}), \qquad {\bm S}({\bm x})={\bm S}_0 \exp(i (\ell-q) {\hat {\bm Q}} \cdot {\bm x})
\end{equation}
where, for the case of interest,  $\phi_0$ and ${\bm S}_0$ can be chosen to be a real amplitude and a real vector, respectively. 
The full free energy density for this ansatz is
\begin{align}
f=& f_{\rm CDW}+f_{\rm SDW}+f_{\rm int}\nonumber\\
=&\frac{1}{2} K_{c} (2\ell-\delta_0)^2 \phi_0^2+\frac{1}{2} \left(r_c+2 {\tilde V}_{\rm eff}(Q_{\rm CDW})\right) \phi_0^2+u_c \phi_0^4
+\frac{\kappa_0}{2} (\Delta {\bar \rho}+2\Lambda |\phi_0|^2 \delta)^2\nonumber\\
+& \frac{1}{2} (r_s+K_s(\ell-q)^2)\; {\bm S}_0^2+u_s \; {\bm S}_0^4
+2 \lambda \phi_0^2 {\bm S}_0^2+\gamma \phi_0^2 {\bm S}_0^2
\label{eq:f-full}
\end{align}
where the second line is the same as Eq.\eqref{eq:f-ICDW-2} (with $\delta=2\ell$). 

\begin{figure}[hbt]
\begin{center}
\includegraphics[width=0.8\textwidth]{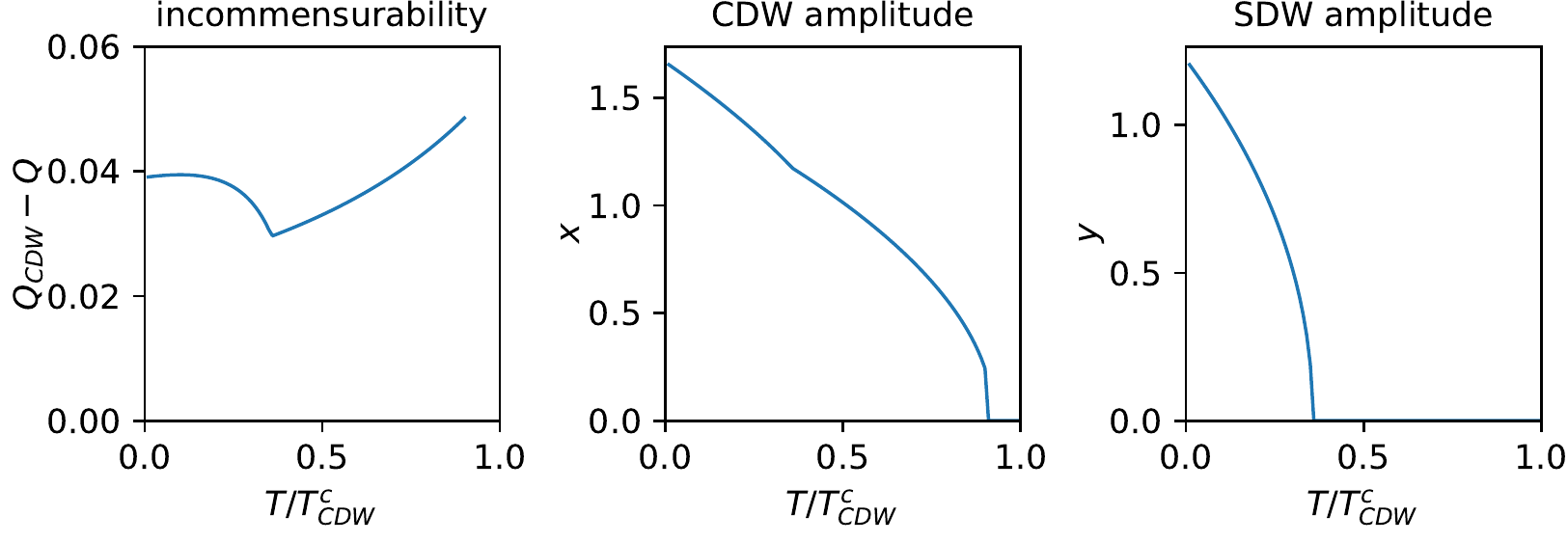}
 \end{center}
 \caption{Incommensurability $\ell$, CDW order parameter $x$ and SDW order parameter $y$ as a function of temperature for typical values of the parameters of the free energy of Eq.\eqref{eq:dimensionless}.}
 \label{fig:non-monotonic}
 \end{figure}

It is convenient to introduce the parametrization
\begin{equation}
x=\frac{\phi_0}{a}, \qquad y=\frac{S_0}{c}, \qquad t=\frac{q}{\delta_0}, \qquad
a^2=\frac{\Delta {\bar \rho}}{2\Lambda \delta_0}, \qquad b^2=\frac{K_c}{4\Lambda^2 \kappa_0}, \qquad z=\frac{b^2}{a^2}, \qquad c^2=\frac{16 \Lambda^2 \kappa_0}{K_s} a^4
\label{eq:parameters}
\end{equation}
Extremizing the free energy of Eq.\eqref{eq:f-full} with respect to $\ell$ we now find
\begin{equation}
\ell=\frac{\delta_0}{2}\Big[\frac{x^2 (z-1)+2ty^2}{x^2(z+x^2)+y^2}\Big]
\label{eq:ellwithq}
\end{equation}
The equilibrium state is now found by finding the minimum of the free energy with respect to $\phi_0$ and $S_0$.  In terms of the parametrization defined in Eq.\eqref{eq:parameters} and the expression for $\ell$ of Eq.\eqref{eq:ellwithq}, the free energy density of Eq.\eqref{eq:f-full} takes the form
\begin{equation}
f=\frac{1}{2}K_c \delta_0^2 a^2 F(x, y)
\label{eq:rescaling}
\end{equation}
where
\begin{align}
F(x,y)=&\Big[\frac{x^2(x^2+1)-(2t-1)y^2}{x^2(x^2+z)+y^2}\Big]^2 x^2+\frac{x^2}{z} \Big[ \frac{z(x^2+1)+2ty^2}{x^2(x^2+z)+y^2}\Big]^2+\mu y^2\Big[\frac{x^2(z-1-2t(x^2+z)}{x^2(x^2+z)+y^2}\Big]^2\nonumber\\
+&{\tilde r}_c x^2+w_cx^4+{\tilde r}_sy^2+w_sy^4+{\tilde \gamma} x^2 y^2+ {\tilde \lambda} xy^2
\label{eq:dimensionless}
\end{align}
Here we used the rescaled parameters
\begin{align}
{\tilde r}_c=&\frac{(r_c+2{\tilde V}(Q)}{K_c \delta_0^2}, \qquad w_c=\frac{2a^2}{K_c \delta_0^2}u_c, \qquad {\tilde r}_s=\frac{c^2}{K_ca^2\delta_0^2}r_s, \qquad 
w_s=\frac{2c^4}{K_c\delta_0^2a} u_s \nonumber\\
{\tilde \gamma}=&\frac{2c^2}{K_c\delta_0^2} \gamma, \qquad {\tilde \lambda}=\frac{4c^2}{K_c\delta_0^2a}\lambda, \qquad \mu=\frac{1}{z}
\label{eq:rescaled_param}
\end{align}
The resulting equilibrium state exhibits the expected non-monotonic behavior of the ordering wave vectors as a function of temperature, showing a kink as the CDW and SDW orders begin to converge to a state close to mutually commensurability (controlled by $q$). An example is shown in Fig.\ref{fig:non-monotonic}.

\subsubsection{Incommensurability fitting}

The temperature dependence of the incommensurability, as plotted in Fig.~\{main text figure\}, can be fit to the model by considering the parameters in Eq.~\eqref{eq:rescaled_param}, minimizing Eq.~\eqref{eq:dimensionless} with respect to the rescaled order parameters $x$ and $y$, and determining the ordering wavevectors via Eq.~\eqref{eq:order_wavevec} and Eq.~\eqref{eq:ellwithq}. Temperature dependence is contained implicitly in the parameters
\begin{equation}
{\tilde r}_c = a_c (T - T_{\rm CDW}^c), \qquad {\tilde r}_s = a_s (T - T_{\rm SDW}^c).
\end{equation}
Here we have introduced the parameters $a_c$ and $a_c$, to convert ${\tilde r}_c$ and ${\tilde r}_s$ into Kelvin. $T_{\rm CDW}^c$ and $T_{\rm SDW}^c$ represent nominal ordering temperatures of CDW and SDW, respectively, i.e. the ordering temperatures in the absence of compressibility effects and coupling between CDW and SDW. The actual ordering temperatures, determined by minimizing Eq.~\eqref{eq:dimensionless}, can vary slightly for the CDW transition and moderately for the SDW transition.

To reduce the number of parameters involved, we fix a few parameters as follows. Noting that the experimental data does not show a clear transition up to $\sim 300\mathrm{K}$, we interpret the charge ordering as having a very high mean-field transition temperature and hence fix $T_{\rm CDW}^c = 400\mathrm{K}$. For the nominal SDW ordering temperature, we fix $T_{\rm SDW}^c = 35\mathrm{K}$ to be roughly consistent with the SDW onset temperature seen in neutron scattering. Finally, the effect of the coupling ${\tilde \gamma}$ can be mimicked by varying the magnitude of ${\tilde \lambda}$. Therefore, ${\tilde \gamma}$ is not strongly constrained and we set it to $0$ for simplicity; the quality of the final fits is not affected.

The remaining parameters $Q, \delta_0, {\tilde \lambda}, z, t, a_c, a_s$ are free. We numerically optimize over the space of these parameters, minimizing $F(x,y)$ and calculating $Q_{\rm CDW}$ as a function of $T$ for each parameter set, to obtain a least-squares fit to the experimental data of incommensurability. The obtained parameters are shown in Table.~\ref{tab:lgfit}. In all cases, the fit is successful, with the statistic $\chi^2 \sim N$, where $N$ is the number of experimental data points.

\begin{table}[hbt]
\centering
\begin{tabular}{|c|c|c|c|c|c|c|c|} 
\hline
$x$ & $Q$ & $\delta_0$ & ${\tilde \lambda}$ & $z$ & $t$ & $a_c$ & $a_s$ \\
\hline
0.10 & 0.0561 & 0.387 & -0.122 & 2.96 & 0.572 & 0.0247 & 0.018 \\
0.125 & 0.106 & 0.293 & -0.899 & 2.65 & 0.523 & 0.0198 & 0.0283 \\
0.15 & 0.132 & 0.197 & -1.32 & 4.1 & 1.23 & 0.0178 & 0.0202 \\
0.17 & 0.246 & 0.114 & -3.58 & 1.08 & 1.16 & 0.0213 & 0.00429 \\
0.20 & 0.23 & 0.0522 & -0.946 & 0.775 & 0.373 & 0.0151 & 0.0562 \\
\hline
\end{tabular}
\caption{Landau-Ginzburg fitting parameters.}
\label{tab:lgfit}
\end{table}

\subsection{Disorder and the Non-linear Sigma Model}

We have not included the effects of quenched disorder in the free energy of Eq.\eqref{eq:LG-CDW}. The local electrostatic potential $V_{\rm dis}({\bm x})$ generated by quenched disorder couples linearly to the local charge density through a term of the form $F_{\rm disorder}=\int d^3 x  V_{\rm dis}({\bm x}) \rho({\bm x})$. Thus, disorder acts a local ``random field'' coupled linearly to the CDW order parameter $\psi({\bm x})$. As it is well known [\cite{imry-1975}], below a critical dimension quenched disorder destroys long range order. In systems with continuous symmetries, such as an ICDW, it completely destroys the true long range order below four dimensions. 

For a basic characterization of the effects of disorder on the structure factor. This problem was discussed extensively in the literature of the Ising model in random fields and related systems [\cite{Kogon-1981,Pytte-1981,Hagen-1983}]. Here we follow the more recent discussion of Ref. [\cite{nie-2014}] and write the structure factor $S$ in terms of an ideal disorder-free susceptibility $G$
\begin{align}
S(Q_{\rm CDW} + q) &= T G(q) + \sigma^2 G^2(q) \label{eq:SqGq}\\
G(q) &= \frac{1}{\kappa_{\parallel} q_x^2 + \kappa_{\perp} q_y^2 + \mu}.
\end{align}
Here, $\sigma$ characterizes the disorder strength, $\kappa_\parallel$ and $\kappa_\perp$ are coefficients of gradient terms in the effective Hamiltonian of [\cite{nie-2014}]. The peak width is determined by the ratio of $\mu$ to $\kappa_\parallel$. The temperature dependence is contained implicitly in $\mu$. After approximating the system as a non-linear sigma model, $\mu$ can be determined self-consistently by the condition
\begin{equation}
1 = \int \frac{d^2 q}{(2\pi)^2} S(q).
\end{equation}
Inserting Eq.~\eqref{eq:SqGq} gives
\begin{equation}
1 = T A_1(\mu) + \sigma^2 A_2(\mu)
\end{equation}
where 
\begin{equation}
A_n(\mu) = \int \frac{d^2 q}{(2\pi)^2} [G(q)]^n.
\end{equation}
Evaluating for $n = 1, 2$, 
\begin{align}
A_2(\mu) &= \frac{1}{4\pi \sqrt{\kappa_{\parallel}\kappa_{\perp}}} \frac{1}{\mu} \\
A_1(\mu) &= \frac{1}{4\pi \sqrt{\kappa_{\parallel}\kappa_{\perp}}} \int_\mu^\Gamma \frac{dx}{x} \\
&= \frac{1}{4\pi \sqrt{\kappa_{\parallel}\kappa_{\perp}}} \ln\frac{\Gamma}{\mu}
\end{align}
where $\Gamma$ is a UV cutoff. Thus, the temperature dependence of $\mu$ is determined by
\begin{equation}
4\pi \sqrt{\kappa_{\parallel}\kappa_{\perp}} = T \ln\left[\frac{\Gamma}{\mu}\right] + \frac{\sigma^2}{\mu}.
\end{equation}
For simplicity, below we will work with $\kappa_\parallel = \kappa_\perp = \kappa$.

\subsubsection{$S(Q)$ lineshape fitting}

The parameters $\sigma^2, \kappa, \Gamma$ make a precise prediction for the evolution of the lineshape of $S(Q)$ as a function of temperature. The applicability of this model can be seen by fitting the entirety of momentum and temperature dependence in the $S(Q, T)$ using Eq.~\eqref{eq:SqGq} with temperature-independent parameters $\sigma^2, \kappa, \Gamma$ and an overall temperature-independent normalization. (Here, since we are only fitting the lineshape, we always consider momentum relative to $Q_{\rm CDW}$). Despite the simplicity of the model, having only four temperature-indepedent parameters, we find that the entire range of temperature in the experimental data, spanning around an order of magnitude, can be fit surprisingly well. The obtained parameters are shown in Table.~\ref{tab:sqtfit}. This suggests that our highly approximate model provides an appropriate description of the effects of disorder on the lineshape of the spectra and its temperature dependence. 

\begin{table}[hbt]
\centering
\begin{tabular}{|c|c|c|c|c|} 
\hline
$x$ & $\sigma$ & $\kappa$ & $\Gamma$ & $A$ \\
\hline
0.10 & 4.28 & 38.4 & 7.49 & 9.65$\times 10^{-4}$ \\
0.125 & 1.56 & 28.1 & 1.42 & 5.13$\times 10^{-4}$ \\
0.15 & 1.10 & 22.3 & 0.663 & 6.08$\times 10^{-4}$ \\
0.17 & 1.13 & 16.4 & 0.362 & 4.46$\times 10^{-4}$ \\
0.20 & 4.41 & 22.9 & 0.782 & 6.30$\times 10^{-4}$ \\
\hline
\end{tabular}
\caption{$S(Q, T)$ fitting parameters. $A$ is an overall normalization factor.}
\label{tab:sqtfit}
\end{table}


\end{document}